\documentclass[journal]{IEEEtran}
\usepackage{graphicx} 
\usepackage{cite}

\usepackage{amsmath,amssymb}
\usepackage{algorithm}
\usepackage{algorithm}
\usepackage[noend]{algpseudocode}
\makeatletter
\newcommand{\LongState}[1]{%
  \State \parbox[t]{\dimexpr\linewidth-\algorithmicindent}{#1}%
}

\algrenewcommand\algorithmiccomment[1]{\hfill$\triangleright$~\parbox[t]{.62\linewidth}{\raggedright #1}}
\makeatletter
\DeclareUnicodeCharacter{2248}{\ensuremath{\approx}} 
\usepackage{booktabs}
\usepackage{multirow}
\usepackage{subcaption}
\usepackage{wrapfig}
\usepackage{float}
\usepackage{hyperref} 

\hypersetup{
    colorlinks=true,
    linkcolor=blue,
    citecolor=blue,
    filecolor=magenta,
    urlcolor=cyan,
}
\ifCLASSINFOpdf
\else
\fi
\begin{document}
%
\title{Network Traffic Classification Using Self-Supervised Learning and Confident Learning}
%
%
%

\author{Ehsan~Eslami,~\IEEEmembership{Student Member,~IEEE,}
        and~Walaa~Hamouda,~\IEEEmembership{Senior Member,~IEEE}
\thanks{E.Eslami and W. Hamouda are with the Department of Electrical and Computer Engineering, Concordia University, Montreal, QC, H3G 1M8, Canada (e-mail: ehsan.eslami@mail.concordia.ca; walaa.hamouda@concordia.ca).}
}

%
%

\markboth{}%
{E. Eslami \MakeLowercase{\textit{et al.}}: Network Traffic Classification Using Self-Supervised Learning and Confident Learning}
%



\maketitle

\begin{abstract}
Network traffic classification (NTC) is vital for efficient network management, security, and performance optimization, particularly with 5G/6G technologies. Traditional methods, such as deep packet inspection (DPI) and port-based identification, struggle with the rise of encrypted traffic and dynamic port allocations. Supervised learning methods provide viable alternatives but rely on large labeled datasets, which are difficult to acquire given the diversity and volume of network traffic. Meanwhile, unsupervised learning methods, while less reliant on labeled data, often exhibit lower accuracy. To address these limitations, we propose a novel framework that first leverages Self-Supervised Learning (SSL) with techniques such as autoencoders or Tabular Contrastive Learning (TabCL) to generate pseudo-labels from extensive unlabeled datasets, addressing the challenge of limited labeled data. We then apply traffic-adopted Confident Learning (CL) to refine these pseudo-labels, enhancing classification precision by mitigating the impact of noise. Our proposed framework offers a generalizable solution that minimizes the need for extensive labeled data while delivering high accuracy. Extensive simulations and evaluations, conducted using three datasets (ISCX VPN-nonVPN, self-generated dataset, and UCDavis--QUIC), and demonstrate that our method achieves superior accuracy compared to state-of-the-art techniques in classifying network traffic.
\end{abstract}


%
\IEEEpeerreviewmaketitle

\section{Introduction}

\IEEEPARstart{N}{etwork} traffic classification (NTC) is a cornerstone of modern network management and security, enabling the identification of applications and services, such as video streaming or data traffic applications, based on their traffic patterns \cite{1}. This capability supports essential functions like Quality of Service (QoS) provisioning \cite{2}, anomaly detection, and traffic engineering. The exponential growth of Internet traffic \cite{3}, driven by the increasing number of connected devices, the expansion of cloud computing, and the rise of Artificial Intelligence (AI)–driven services, highlights the need for effective NTC methods \cite{4}. The deployment of 5G and upcoming 6G networks has further complicated network traffic \cite{5}, introducing diverse requirements for network slicing, edge computing, and Ultra-Reliable Low-Latency Communications (URLLC), making robust classification increasingly essential \cite{1,6}. AI plays a vital role in modern networking, facilitating intelligent traffic management, predictive maintenance, and real-time anomaly detection \cite{7,8}. 

For NTC, traditional approaches such as port-based identification and Deep Packet Inspection (DPI) have become inadequate. Port-based methods fail due to dynamic port usage, while DPI struggles with the prevalence of encrypted traffic \cite{2,9,10}. As a result, modern supervised and representation-learning approaches now dominate \cite{2}. Machine Learning (ML) and Deep Learning (DL) methods can classify traffic using statistical features \cite{11} and time-series features \cite{12}. Supervised learning trains models on labeled datasets and can achieve state-of-the-art (SOTA) performance when sufficient labels are available. For instance, classical supervised models such as Support Vector Machines (SVM) and logistic regression remain competitive in Software-Defined Networking (SDN) settings \cite{13}; information-bottleneck neural networks report strong accuracy and robustness on common benchmarks \cite{14}; supervised contrastive variants achieve leading results under label imbalance \cite{15}; and robust supervised architectures such as RBLJAN, which couple byte–label joint attention with adversarial training, sustain accuracy under perturbations \cite{16}. Unsupervised learning, in contrast, does not require labels and can discover patterns independently \cite{1}. More recently, Self-Supervised Learning (SSL) has shown promise by leveraging abundant unlabeled traffic to learn representations and produce pseudo-labels with a small labeled set, reducing reliance on extensive labeled datasets and improving coverage \cite{17}. For example, contrastive pretraining on flow images achieves strong few-shot and transfer performance \cite{18}; transformer pretraining, such as ET-BERT tokenizes packet/flow bytes and transfers to downstream NTC tasks \cite{19}; and masked autoencoding of flow images in YaTC advances SOTA SSL representations \cite{20}. SSL tailored to encrypted protocols likewise reports high accuracy with only a small labeled seed \cite{21}.

Despite this progress, important challenges remain. For supervised learning, labeling network traffic is resource-intensive and costly, and labeled datasets often fail to reflect the full diversity of real-world traffic, which reduces model generalizability \cite{22}, \cite{23}, \cite{24}. Recent dataset-forensics show that capture-specific fields (e.g., timestamps/MAC/IP/ports) in common NTC benchmarks inflate supervised results; once properly anonymized, accuracy drops sharply, exposing limited generalization \cite{25}. Unsupervised methods, while label-free, often struggle to capture the complex patterns needed for precise traffic classification and typically underperform supervised methods \cite{2,4}. While SSL offers significant benefits by leveraging unlabeled data, SSL-derived pseudo-labels can contain noise due to imperfect pretext accuracy; if used directly to train the downstream classifier, these errors degrade performance. A principled way to cope with noisy labels is confident learning (CL) \cite{26}, which estimates class-conditional label errors from out-of-sample predicted probabilities together with the observed (pseudo) labels and constructs the confident joint—a matrix summarizing how often each observed class appears to belong to each (latent) true class—thus enabling noise-aware filtering or reweighting of pseudo-labels. Bringing CL principles to NTC addresses pseudo-label noise directly and provides a principled bridge from representation learning to reliable training.

To address the aforementioned challenges, this paper proposes a novel framework that combines SSL and CL for NTC. First, our approach uses SSL to generate pseudo-labels for unlabeled traffic flows by pretraining with an unlabeled dataset and fine-tuning with a small labeled dataset. We develop two complementary SSL branches—(i) an autoencoder (AE) with constraint-consistent reconstruction for heterogeneous (continuous/categorical) flow features and (ii) Tabular Contrastive Learning (TabCL) \cite{27} with constraint-preserving projection and dual-head projections with head-specific temperatures. We then fuse the SSL predictions via a confidence/margin voting rule. Second, inspired by CL, we decouple noise estimation from the final classifier and treat the SSL pseudo-labels as inputs to a CL module. Using $k$-fold out-of-sample prediction probabilities, we estimate the confident joint and introduce adaptations tailored to network traffic flows: per-class, quantile-calibrated thresholds (rather than means) that adapt to class imbalance and calibration drift; a calibration-aware logistic weighting that smoothly down-weights uncertain pseudo-labels instead of discarding data; and a balanced retention constraint that aligns each class’s retained training mass with its CL-estimated clean fraction \cite{26}. Third, we train the downstream classifier on the full pseudo-labeled pool using a weighted symmetric cross-entropy objective, where per-sample weights—derived from the CL module via out-of-sample probabilities—down-weight uncertain pseudo-labels while retaining all data.

Although recent SOTA methods attain high accuracy, important gaps persist—namely, reliance on scarce labels in supervised learning, limited precision in unsupervised methods, noise in SSL pseudo-labels, and per-class calibration/imbalance issues. Our framework addresses these gaps by pairing NTC-specific SSL (constraint-aware views, dual-head temperatures, SSL fusion) with NTC-adapted CL (quantile thresholds, logistic weighting, balanced retention) prior to final training. The contributions of this paper are as follows:

\begin{enumerate}
    \item We propose a novel framework integrating SSL and CL for NTC that first produces pseudo-labels via AE- and TabCL-based SSL, then denoises them with CL before final training.
    
    \item We demonstrate the effectiveness of AE with constraint-consistent reconstruction and TabCL with class-conditioned, constraint-preserving views and dual-head projections for generating high-quality pseudo-labels from unlabeled data, enhancing the utility of SSL.
    
    \item We show that CL improves classification accuracy by identifying and attenuating noisy pseudo-labels using per-class quantile thresholds, logistic weights for smooth attenuation, and balanced retention guided by the confident joint.

    \item We conduct a thorough evaluation, including simulations to create a realistic network traffic dataset, and assess our model’s performance by comparing it to existing methods.

\end{enumerate}

The rest of this paper is organized as follows: Section \ref{sec:related_work} reviews related works in NTC and SSL. Section~\ref{sec:problem_formultion} formalizes the problem, assumptions, and notation. Section \ref{sec:centralized_approach} details the proposed framework. Section \ref{sec:results} presents the simulation and experimental results, and Section \ref{sec:conclusion} concludes the paper.

\section{Related Works}
\label{sec:related_work} 
This section provides a detailed review of the SOTA research, focusing on ML, supervised and unsupervised DL, and SSL methods.

\subsection{Traditional and Machine Learning for NTC}
Early NTC relied on port-based methods and DPI. \cite{28} propose BitProb, which encodes each flow as an $n$-bit signature and uses a Probabilistic Counting Deterministic Automaton (PCDA) to assign the most likely protocol class. These traditional methods struggle with dynamic port and encrypted traffic \cite{4}. ML is one of the promising solutions for NTC \cite{13}, \cite{29}, \cite{30}. For example, \cite{29} utilized supervised ML models such as Decision Tree and Random Forest in an SDN context for NTC, achieving high accuracy. R. H. Serag et al. \cite{13} proposed an ML-based traffic classification method for SDNs, demonstrating the effectiveness of classical ML algorithms like SVM and Logistic Regression in modern network architectures. These approaches handle encrypted and dynamic traffic well using statistical and flow features, but remain constrained by the high labeling demand of supervised learning.

\subsection{Supervised and Unsupervised Deep Learning for NTC}
DL has shown promise in NTC due to its ability to learn complex patterns from raw data \cite{14}, \cite{15}, \cite{16}, \cite{31}, \cite{32}, \cite{33}, \cite{34}, \cite{35}. For instance, Lotfollahi et al. \cite{31} proposed “Deep Packet,” using Stacked Autoencoders (SAEs) and Convolutional Neural Networks (CNNs) to classify encrypted traffic, achieving 98\% accuracy on the ISCX VPN-nonVPN dataset \cite{12}, while reducing manual feature engineering via learned representations. \cite{32} introduced a DL approach combining CNN and Recurrent Neural Networks (RNNs), achieving SOTA performance on the  CIC-IOT dataset \cite{36}. In  \cite{14}, IBNN adds an information-bottleneck layer to a CNN plus a fully connected classifier to retain task-relevant features while discarding noise, achieving 97.6\%/98.2\% on ISCX datasets; ablation without the IB layer drops ~8\%, indicating stronger generalization and robustness. Ma et al. \cite{15} introduce Balanced Supervised Contrastive Learning (BSCL), which maps the first 25 packets to compact traffic images and trains a ResNet-50 with supervised contrastive learning plus imbalance-aware losses; On the ISCX2016 it achieves high traffic-type accuracy and near-perfect per-class results. \cite{33} proposes a CNN–LSTM (Long-Short-Term Memory) autoencoder that learns spatiotemporal flow features from statistics and then classifies in latent space, reaching near-perfect accuracy on 1.42M flows (4 classes) while remaining robust to imbalance. Xiao et al.’s RBLJAN \cite{16} is a supervised DL model that processes headers and payloads in parallel and applies byte–label joint attention, with training by a GAN-based adversarial traffic generator; it uses focal loss + label regularization and extends from packet to flow level via a dedicated architecture. On various datasets, RBLJAN reports 98\% average F1 and about 97\% accuracy. MTEFU \cite{35} uses multi-task supervision (class + bandwidth + duration) with shared CNN, SAE, Gated Recurrent Unit (GRU), and LSTM backbones with early-packet inputs to boost label efficiency, reaching 94.00\% class accuracy on QUIC with 150 labels. Despite their success, these supervised DL methods rely heavily on large labeled datasets, which are costly and may not generalize across diverse network environments.

On the other hand, unsupervised learning methods have been explored for NTC to circumvent the need for labeled datasets \cite{37}, \cite{38}, \cite{39}. Zhang et al. \cite{37} proposed an adversarial training and deep clustering method. Their approach achieved a multi-classification accuracy of 92.2\% on the USTC-TFC2016 dataset. \cite{38} propose an unsupervised deep multi-source domain-adaptation NTC model (multi-scale features + adversarial/marginal/conditional alignment + weighted, consistency-calibrated classifier) that achieves ~89\% accuracy. These methods cluster flows by similarity without labels; despite this advantage, they generally lag behind supervised learning in accuracy, particularly in complex and diverse traffic scenarios.

\subsection{Self-Supervised Learning for NTC}
By learning directly from unlabeled data, SSL methods reduce reliance on large labeled datasets  \cite{18}, \cite{19}, \cite{20}, \cite{21}, \cite{40}, \cite{41}, \cite{42}. Towhid and Shahriar \cite{21} applied SSL techniques to encrypted NTC,
achieving high accuracy with minimal labeled data across multiple public datasets such as Quick UDP Internet Connection (QUIC) \cite{43} and ISCX VPN-nonVPN datasets. \cite{18} proposes an SSL framework for traffic classification that utilizes the FlowPic method. This approach transforms network flow data into 2D histogram images, which are then processed by contrastive representation learning and CNNs. ET-BERT \cite{19} is a Transformer-based framework for NTC that learns contextualized datagram representations from large unlabeled traffic and then fine-tunes with few labels. It uses Datagram2Token (BURST extraction + byte-bigram tokenization) and two traffic-tailored pretext tasks—Masked BURST Modeling (MBM) and Same-origin BURST Prediction (SBP)—with a 12-layer bidirectional Transformer, supporting packet- and flow-level fine-tuning. ET-BERT delivers strong macro-F1 and accuracy ($>96\%$) on different datasets. YaTC \cite{20} couples a masked-autoencoder (MAE) plus packet/flow-level Transformer attention over a multi-level flow representation (MFR) byte layout. Pretrained on large unlabeled traffic and fine-tuned with few labels, YaTC achieves SOTA on multiple NTC datasets. These advancements highlight SSL’s potential to reduce dependency on labeled data, making it a key focus of our work. 

\subsection{Comparison with Existing Methods}
We integrate constraint-aware SSL methods for pseudo-labeling, a traffic-adapted CL module to denoise them, and a weighted Multilayer Perceptron (MLP) for final training. This differs from supervised NTC methods \cite{13, 14, 15, 16}, \cite{29}, \cite{31, 32, 33, 34, 35}, which require extensive labels, by leveraging unlabeled traffic while controlling label noise. Our constraint-consistent-reconstruction AE and constraint-aware TabCL with dual-head temperatures, combined with traffic-adopted CL extensions, yield better scalability and accuracy than prior SSL baselines \cite{18, 19, 20, 21}, \cite{40, 41, 42}.

\section{Problem Definition and Formulation}
\label{sec:problem_formultion}
We consider NTC over $K$ application classes from heterogeneous flow features. Let $D_s=\{(\mathbf{x}_i,y_i)\}_{i=1}^{|D_s|}$ denote a small labeled dataset with $y_i\in\{1,\dots,K\}$ and $D_l=\{\mathbf{x}_j\}_{j=1}^{|D_l|}$ a large unlabeled dataset drawn from the same deployment domain. Each flow $\mathbf{x}\in\mathbb{R}^{d}$ comprises continuous attributes $\mathbf{x}_{\text{cont}}$ (e.g., counts, bytes, durations, rates) and categorical attributes $\mathbf{x}_{\text{cat}}$ (e.g., protocol, direction), encoded numerically. Continuous attributes include statistical features (e.g., total packets, total bytes, flow duration, average packet size) and time-series features (e.g., inter‐arrival times, burst patterns). These features are essential for classification because they capture both the aggregate behavior and temporal dynamics of traffic. 

Given $D_s$ and $D_l$, our objective is to learn a multiclass classifier $\mathcal{C}_{\text{final}}:\mathbb{R}^{d}\!\to\!\{1,\dots,K\}$ that attains high accuracy and macro-F1 under scarce labels by leveraging $D_l$ via self-supervision and training on a noise-aware pseudo-labeled pool. Concretely, we first generate pseudo-labels for $D_l$ through two complementary SSL branches adapted to NTC: (i) an AE with constraint-consistent reconstruction for heterogeneous flow features, and (ii) TabCL with class-conditioned, constraint-preserving augmentations and dual-head projections for feature-type heterogeneity. When both branches are used, their predictions are fused by a simple confidence/margin voting rule. Because such labels may be noisy, we explicitly account for label noise via the CL module tailored to NTC via per-class quantile thresholds, calibration-aware logistic weights, and balanced retention. We then train $\mathcal{C}_{\text{final}}$ on the resulting pool with confidence-aware reweighting.

We adopt a closed-world setting with $K$ known classes. Real-world NTC exhibits class imbalance and calibration drift across classes; pseudo-labels produced by SSL contain label noise. We assume access to flow/header-derived features (no deep packet inspection of payloads). Our focus is robustness to label scarcity and pseudo-label noise. To maintain clarity and consistency, we define the notations used in sections~\ref{sec:problem_formultion}--\ref{sec:centralized_approach} in Table \ref{tab:notation}.

\begin{table}[t]
    \centering
    \caption{List of Notations.}
    \label{tab:notation}
    \begin{tabular}{c|p{5.2cm}}
        \hline
        \textbf{Symbol} & \textbf{Description} \\
        \hline
        $\mathbf{x}$ & Flow feature vector in $\mathbb{R}^{d}$ \\
        $\mathbf{x}_{\text{cont}}$, $\mathbf{x}_{\text{cat}}$ & Continuous / categorical feature slices \\
        $y$, $\tilde{y}$ & True label, pseudo-label \\
        $D_s$ & Small labeled dataset $\{(\mathbf{x}_i,y_i)\}$ \\
        $D_l$ & Large unlabeled dataset $\{\mathbf{x}_i\}$ \\
        $K$ & Number of application classes \\
        $d$ & Feature dimension \\
        $\mathcal{E}$, $\mathcal{D}$, $\mathcal{C}$ & Encoder, decoder, classifier \\
        $\widehat{\mathbf{x}}$ & Autoencoder reconstruction of $\mathbf{x}$ \\
        $\mathcal{L}$, $\theta$ & Loss function, model parameters \\
        $\mathcal{G}$ & Set of algebraic constraints \\
        $g$ & residual function \\
        $\phi$ & Weights for constraint residuals in AE \\
        $\widetilde{\mathbf{x}}^{(t)}_i$ & replaced view in TabCL \\
        $\breve{\mathbf{x}}^{(t)}_i$ & Adjusted view by constraint-projection \\
        $r$ & TabCL replacement rate \\
        $\mathcal{P}$ & Projection head (TabCL) \\
        $\mathbf{v}^{(t)}_{i}$ & Contrastive projection vector \\
        $\tau$ & Temperature (TabCL) \\
        $N$ & Mini-batch size \\
        $\lambda$ & Mixing coefficient in final TabCL loss \\
        $F$ & Number of folds for CV  \\
        $\eta$ & Learning rate \\
        $\mathbf{p}$ & Predicted probability vector \\
        $\widehat{c}(\mathbf{x})$ & fused pseudo-label (final vote) \\
        $m(\mathbf{x})$ & Confidence margins (top-1 minus runner-up) \\
        $s_i$ & Self-confidence for sample $i$ \\
        $\mathcal{S}_j$ & Self-confidence set for class $j$ \\
        $t^{(q)}_j$, $q$ & $q$-quantile threshold for class $j$; quantile level \\
        $\sigma_j$ & robust scale for class $j$ \\
        $\gamma$ & slope \\
        $w_i$ & Calibration-aware per-sample weight \\
        $w_{\min}$ & minimum weight for CL \\
        $\widehat{Q}$ & Confident-joint matrix \\
        $\rho_j$ & estimated clean fraction for class $j$ \\
        $n_j$ & Count in class $j$ \\
        $T_j$, $M_j$, $a_j$ &  Target mass; current mass; class scaling factor \\
        $w_i'$ & Final per-sample weight \\
        $\operatorname{clip}(x,\ell,u)$ & Clipping operator $\min(\max(x,\ell),\,u)$ \\
        $\alpha,\beta$ & SCE loss coefficients (forward/reverse CE) \\
        \hline
    \end{tabular}
\end{table}

\section{Self-Supervised Learning for Pseudo-Labeling and Confident Learning}
\label{sec:centralized_approach} 
This section first presents a system-level model and data flow, from raw packet capture to the unlabeled flow dataset and ultimately to classification. We then detail the tailored SSL techniques (AE and TabCL), pseudo-label fusion, and the traffic-adapted confident learning used to train the final classifier.

\subsection{System Model and End-to-End Data Flow} 
Raw packet traces are passively captured at monitored links. Packets are aggregated into bidirectional flows using the 5-tuple (source IP, destination IP, source port, destination port, protocol) with idle/active timeouts; time windows may segment long connections for memory safety. The captured flows are then preprocessed to extract features. These features can be broadly categorized into statistical \cite{11} and time-series features \cite{12}. Statistical features include metrics such as the total number of packets in a flow, the total bytes transferred, the duration of the flow, and the average packet size. These features provide a summary of the flow’s characteristics and can help distinguish between different types of applications. For example, video streaming applications might have larger total bytes transferred compared to text-based chat applications. Time-series features, on the other hand, capture the temporal patterns in the traffic. These can include the inter-arrival times between packets, the distribution of packet sizes over time, or the presence of bursty behavior. Such features are particularly useful for identifying applications with specific timing requirements, such as real-time communications or periodic updates. 

\begin{figure*}[!t]
    \centering
    \includegraphics[width=0.95\linewidth]{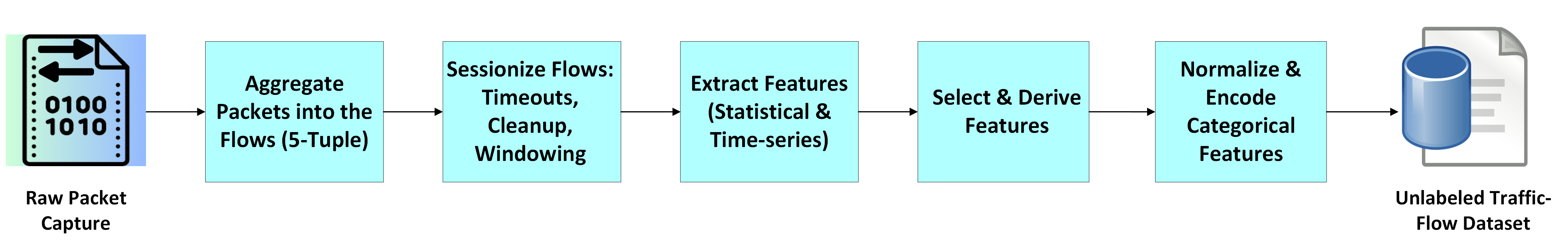}
    \caption{Pipeline from raw packet traces to an unlabeled traffic-flow feature dataset.}
    \label{fig:pipe_line_raw_packet}
\end{figure*}

Beyond the primary features, we also derive secondary features by simple aggregations and ratios (e.g., mean/variance of IAT, byte/packet rates, up/down throughput, and burst-length ratios), which enrich the tabular representation without accessing payloads. In addition to these numeric attributes, there are categorical features such as protocol and flow direction. The categorical fields are encoded as one-hot or embeddings, while numeric fields are standardized/normalized. This produces a large \emph{unlabeled} traffic-flow dataset $D_l$; a small \emph{labeled} subset $D_s$ is curated via DPI/heuristics/manual checks and used only for fine-tuning and model selection. The pipeline of unlabeled dataset generation from raw network packets is illustrated in Fig.~\ref{fig:pipe_line_raw_packet}.  

\begin{figure*}[!t]
    \centering
    \includegraphics[width=0.95\linewidth]{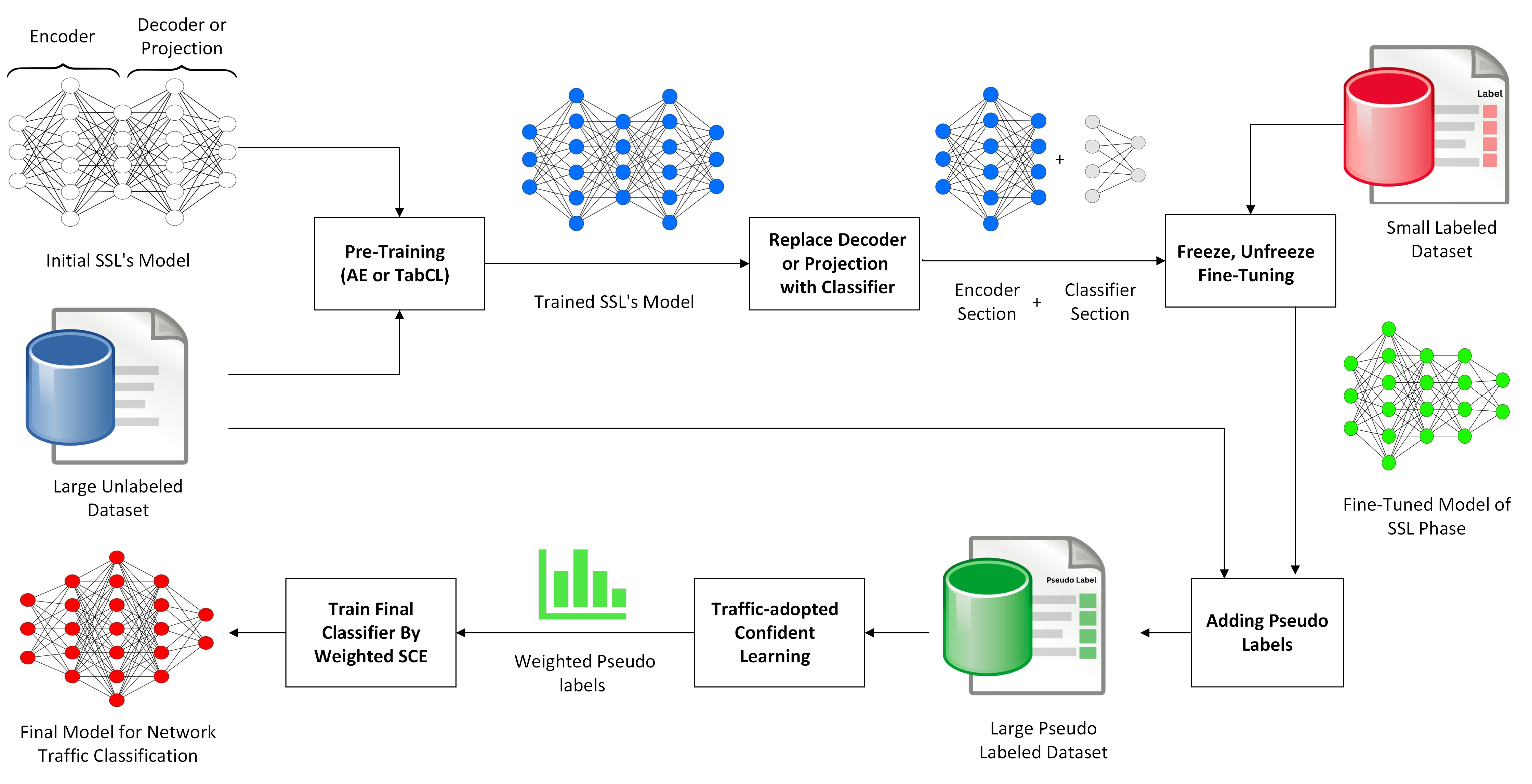}
    \caption{Architectural components of traffic classification with SSL and confident learning.}
    \label{fig:Central_Architecture}
\end{figure*}

As shown in Fig.~\ref{fig:Central_Architecture}, the learning stack has two  SSL methods that operate on $D_l$: an AE tailored to tabular flows, which reconstructs heterogeneous continuous/categorical features; and a TabCL that performs contrastive pretraining with constraint-preserving view generation and dual-head projections specialized for continuous vs.\ categorical slices. After SSL pretraining, the decoder/projection head is replaced by a classifier, and the encoder, along with the classifier, is fine-tuned on the small labeled set $D_s$ (freeze–unfreeze). The fine-tuned encoder(s) are then applied to $D_l$ to produce pseudo-labels; when both methods are used, per-flow predictions are fused via a simple confidence/margin voting rule.

A traffic-adapted Confident Learning (CL) module estimates label noise on the pseudo-labeled pool using out-of-sample probabilities, builds the confident joint, and assigns per-sample training weights via per-class quantile thresholds with calibration-aware smoothing and balanced retention. The final classifier $\mathcal{C}_{\text{final}}$ is trained on the full pseudo-labeled set using weighted symmetric cross-entropy. This overview sets the stage; detailed algorithms and mathematical formulas for AE, TabCL, fusion, and CL follow in the next subsections.

\subsection{Autoencoder-Based Self-Supervised Learning}
An Autoencoder learns a compressed representation of the input data by encoding it into a lower-dimensional latent space and reconstructing it, capturing underlying patterns in an unsupervised manner. We represent each network flow as a feature vector $\mathbf{x}\in\mathbb{R}^{d}$ comprising two types of attributes that are typical of network-traffic data: (i) continuous, numeric features (e.g., counts, durations, or rates) denoted by $\mathbf{x}_{\text{cont}}$, and (ii) categorical features (e.g., protocol, flow direction) denoted collectively by $\mathbf{x}_{\text{cat}}$. Categorical attributes are encoded in standard numeric form (e.g., one-hot or embeddings) and are reconstructed with a cross-entropy objective. This heterogeneous structure is characteristic of tabular network-flow representations and motivates losses that respect both numeric reconstruction fidelity and categorical consistency. We utilize an autoencoder to first learn latent representations from $D_l$ and second, generate pseudo-labels, subsequently fine-tuning with $D_s$.

\subsubsection{Pretraining objective with constraint-consistent reconstruction}
Let $\mathcal{E}(\,\mathbf{x}\,;\theta_{\mathcal{E}})$ be the encoder and $\mathcal{D}(\,\mathbf{h},;\theta_{\mathcal{D}})$ the decoder. Given an input $\mathbf{x}$, the model produces a latent code $\mathbf{h}=\mathcal{E}(\mathbf{x})$ and a reconstruction $\widehat{\mathbf{x}}=\mathcal{D}(\mathbf{h})$, with corresponding splits $\widehat{\mathbf{x}}_{\text{cont}}$ and $\widehat{\mathbf{x}}_{\text{cat}}$. The pretraining loss function combines a mixed reconstruction term with a small, domain-consistent penalty that softly enforces simple algebraic relations among selected continuous coordinates:
\begin{equation}
\label{eq:ae_loss_total}
\mathcal{L}_{\text{AE}} \;=\;
\underbrace{\mathrm{MSE}\!\big(\mathbf{x}_{\text{cont}},\widehat{\mathbf{x}}_{\text{cont}}\big)
+\mathrm{CE}_{\text{cat}}\!\big(\mathbf{x}_{\text{cat}},\widehat{\mathbf{x}}_{\text{cat}}\big)}_{\text{mixed reconstruction}}
\;+\;
\underbrace{\mathcal{L}_{\text{cons}}}_{\substack{\text{constraint}\\\text{consistency}}},
\end{equation}
where $\text{MSE}$ is the mean squared error on continuous features, and $\mathrm{CE}$ denotes the sum of cross-entropy terms over all categorical fields (each field modeled with an appropriate softmax). In network-traffic data, several continuous attributes are related by simple algebra (e.g., a rate-like feature equals a bytes-like feature over a duration-like feature; a packet-length-like feature equals the sum of payload- and header-like features). When the autoencoder reconstructs such tuples, encouraging these relations to hold improves plausibility and stabilizes the latent representation. We formalize this with a set of algebraic \emph{residuals} applied to the reconstructed continuous vector $\widehat{\mathbf{x}}_{\text{cont}}$:
\[
\mathcal{G}\;=\;\big\{\,g_m:\mathbb{R}^{d_{\text{cont}}}\!\to\mathbb{R}\ \mid\ m=1,\dots,|\mathcal{G}|\,\big\},
\]
where each $g_m(\widehat{\mathbf{x}}_{\text{cont}})$ is the residual of constraint $m$ when evaluated on the reconstructed continuous vector. We then define
\begin{equation}
\label{eq:ae_consistency_loss}
\mathcal{L}_{\text{cons}}
\;=\;
\sum_{m=1}^{|\mathcal{G}|}\phi_m\,\big\lVert\, g_m\!\big(\widehat{\mathbf{x}}_{\text{cont}}\big)\,\big\rVert_{1},
\end{equation}
where $\phi_m\!>\!0$ are modest weights. This penalty nudges the encoder and decoder to reconstruct plausible tuples without prescribing any particular feature identity, and improves the semantic quality of the representation reused at fine-tuning time.

\subsubsection{Fine-tuning and pseudo-labeling}
After pretraining on $D_l$ with~\eqref{eq:ae_loss_total}, the decoder is replaced by a classifier $\mathcal{C}(\,\cdot\,;\theta_{\mathcal{C}})$ attached to the encoder. We adopt a short freeze--unfreeze schedule. This schedule is a standard practice that reduces optimization drift when transplanting a new head on a pretrained encoder. First, we freeze $\theta_{\mathcal{E}}$ and optimize $\theta_{\mathcal{C}}$ on $D_s$ to stabilize the new head, then jointly unfreeze $(\theta_{\mathcal{E}},\theta_{\mathcal{C}})$ and fine-tune on $D_s$ with the standard cross-entropy:
\begin{equation}
\label{for:loss_ce}
\mathcal{L}_{\text{CE}}
\;=\;
-\frac{1}{|D_s|}
\sum_{(\mathbf{x},y)\in D_s}
\log \Big(\,\mathcal{C}\big(\mathcal{E}(\mathbf{x})\big)[y]\,\Big).
\end{equation}
After fine-tuning, we apply the trained head to all $\mathbf{x}\in D_l$ and assign pseudo-labels by top-1 prediction.

\subsubsection{Hyperparameter selection}
Latent width, learning rates, the freeze duration, and the coefficients $\{\phi_m\}$ in~\eqref{eq:ae_consistency_loss} are chosen by stratified cross-validation on $D_s$, targeting macro-F1 after fine-tuning. This keeps the added constraint term modest and data-adaptive, avoiding over-regularization on classes whose statistics naturally vary. Algorithm \ref{alg:ae_ssl_cons} summarizes the procedure.

\begin{algorithm}[t]
\caption{Autoencoder-SSL with Constraint-Consistent Reconstruction}
\label{alg:ae_ssl_cons}
\begin{algorithmic}[1]
\Require Unlabeled $D_l$, labeled $D_s$, batch size $m$, learning rate $\eta$
\State Initialize encoder $\mathcal{E}(\cdot;\theta_{\mathcal{E}})$, decoder $\mathcal{D}(\cdot;\theta_{\mathcal{D}})$
\State \textbf{Pretrain on $D_l$:}
\For{epoch $=1,2,\dots$}
  \For{mini-batch $\{\mathbf{x}_i\}_{i=1}^{m}\subset D_l$}
    \State $\mathbf{z}_i\gets \mathcal{E}(\mathbf{x}_i)$; \quad $\widehat{\mathbf{x}}_i\gets \mathcal{D}(\mathbf{z}_i)$
    \LongState{Compute $\mathcal{L}_{\mathrm{AE}}$ from \eqref{eq:ae_loss_total} with $\mathcal{L}_{\mathrm{cons}}$ in \eqref{eq:ae_consistency_loss} (weights $\phi_m$); update $(\theta_{\mathcal{E}},\theta_{\mathcal{D}})\leftarrow(\theta_{\mathcal{E}},\theta_{\mathcal{D}})-\eta\nabla\mathcal{L}_{\mathrm{AE}}$.}
  \EndFor
\EndFor
\State Replace $\mathcal{D}$ by classifier $\mathcal{C}(\cdot;\theta_{\mathcal{C}})$
\State \textbf{Freeze} $\theta_{\mathcal{E}}$; train $\theta_{\mathcal{C}}$ on $D_s$ using CE \eqref{for:loss_ce}
\State \textbf{Unfreeze} $(\theta_{\mathcal{E}},\theta_{\mathcal{C}})$; joint fine-tuning on $D_s$ with \eqref{for:loss_ce}
\State \textbf{Pseudo-labeling}: for all $\mathbf{x}\in D_l$, set $\widehat{c}\gets \arg\max_k \mathcal{C}(\mathcal{E}(\mathbf{x}))[k]$
\Ensure Pseudo-labeled $D_l=\{(\mathbf{x},\widehat{c})\}$
\end{algorithmic}
\end{algorithm}

\subsection{Tabular Contrastive Learning (TabCL)}
\begin{figure}[h]
    \centering
    \includegraphics[width=0.9\columnwidth]{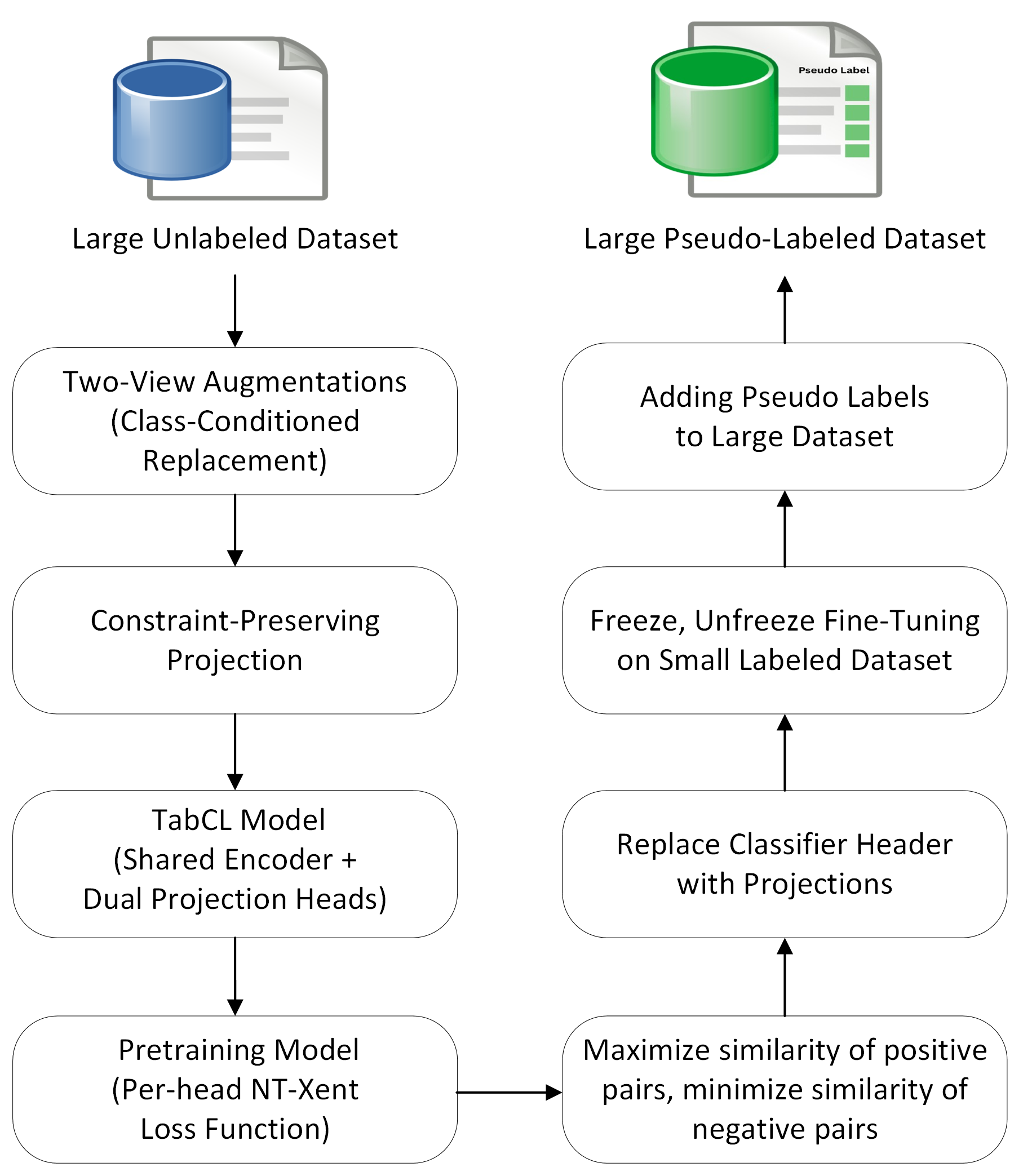}
    \caption{TabCL process for pseudo-labeling.}
    \label{fig:TabCL_FlowChart}
\end{figure}

TabCL offers an alternative SSL approach by learning invariant representations through contrastive loss, enhancing the model’s ability to generalize across diverse traffic flows. TabCL leverages contrastive learning to maximize the similarity between augmented versions of the same data sample (positive pairs) while distinguishing it from others (negative pairs), making it particularly effective for tabular data. The detailed workflow of TabCL method is presented in Fig. \ref{fig:TabCL_FlowChart}. We adopt the standard two-view NT-Xent setup and tailor it to network-traffic data with two components: a class-conditioned replacement scheme to create views, and a constraint-preserving projection that enforces simple identities among reconstructed continuous attributes in network traffic datasets. In addition, we use a dual-head contrastive objective to accommodate heterogeneous feature types with distinct temperature scales.

\subsubsection{Two-view augmentations for tabular flows}
Before pretraining in TabCL, we first obtain primary pseudo-labels $\{\tilde y_i^{(0)}\}$ for $D_l$ by training a light classifier on $D_s$ and applying it to $D_l$. During contrastive pretraining, we periodically refresh these pseudo labels: at predefined intervals, we freeze the current encoder $\mathcal{E}$, train a fresh linear probe on $\{\mathcal{E}(\mathbf{x}),y\}_{(\mathbf{x},y)\in D_s}$, and reclassify $\{\mathcal{E}(\mathbf{x})\}_{\mathbf{x}\in D_l}$ to obtain $\{\tilde y_i^{(t)}\}$. Refreshing stops after a small number of rounds or once the fraction of changed labels falls below a tolerance.

For each anchor $\mathbf{x}_i$ with current class identity $c_i=\tilde y_i^{(t)}$, we create two views $\widetilde{\mathbf{x}}^{(1)}_i$ and $\widetilde{\mathbf{x}}^{(2)}_i$ by replacing a small subset of the anchor’s features. Let $d$ denote the number of features (coordinates) of $\mathbf{x}_i$, and let $r\!\in(0,1)$ be the replacement rate. In each view, we randomly select exactly $\lceil r d\rceil$ features of the anchor to replace. Denote by $\mathcal{I}^{(t)}\subset\{1,\dots,d\}$ the index set chosen for view $t\in\{1,2\}$; the two sets are sampled independently (preferably with minimal overlap). For every index $j\in\mathcal{I}^{(t)}$, the new value is drawn from the empirical distribution of the same feature among flows currently labeled $c_i$, while all other features are kept equal to the anchor:
\begin{align}
\widetilde{\mathbf{x}}^{(t)}_i[j] \;&\sim\; \mathbb{P}\!\big(\mathbf{x}[j]\mid \tilde y=c_i\big), 
&& j\in\mathcal{I}^{(t)},\\[-2pt]
\widetilde{\mathbf{x}}^{(t)}_i[k] \;&=\; \mathbf{x}_i[k], 
&& k\notin\mathcal{I}^{(t)}.
\end{align}
This class-conditioning replacement (TabCL principle) keeps views semantically close to the anchor $\mathbf{x}_i$ while inducing invariances within the class manifold.

After replacement, we update each view to satisfy simple algebraic identities among continuous features, using the residual set $\mathcal{G}$ defined in the Autoencoder subsection. In practice, we adjust the augmented vector only as much as needed to restore identities (e.g., rate = bytes / duration” or “packet length = payload + header), keeping the view close to the original replacement outcome while making the tuple plausible in terms of inter-feature consistency. Let $\breve{\mathbf{x}}^{(t)}_i$ denote the adjusted view:
\begin{align}
\label{eq:tabcl_proj}
\breve{\mathbf{x}}^{(t)}_i \;=&\; \Pi_{\mathcal{G}}\!\big(\widetilde{\mathbf{x}}^{(t)}_i\big), \\[2pt]
\triangleq\;&\; \arg\min_{\mathbf{u}}\; \big\lVert \mathbf{u}-\widetilde{\mathbf{x}}^{(t)}_i\big\rVert_2^2 \nonumber\\[-2pt]
&\text{s.t.}\;\; g_m\!\big(\mathbf{u}_{\text{cont}}\big)=0,\ \forall g_m\in\mathcal{G}. \nonumber
\end{align}
Here, $\mathbf{u}$ is the optimization variable (the adjusted vector we solve for), $\mathbf{u}_\text{cont}$ is its continuous slice, $\Pi_\mathcal{G} (.)$ denotes the projection operator that returns the nearest vector that satisfies all constraints in $\mathcal{G}$, and $\arg\min$ selects the point that minimizes the squared distance to the replaced view.
This constraint-preserving update is simple (closed-form or single-pass updates for ratio/product/sum identities) and ensures that positives are plausible flow tuples. Although applied at the view level, the induced gradients during pretraining bias the encoder’s representation toward respecting such identities, which we later reuse for supervised fine-tuning.

\subsubsection{Dual-Head Projections for heterogeneous features}
After creating two augmented views by class-conditioning replacement and constraint-preserving update, in the pretraining phase, we adopt the Normalized Temperature-scaled Cross-Entropy (NT-Xent) loss promotes similarity between augmented views. For projection vectors $\mathbf{v}_i,\mathbf{v}_j$ and temperature $\tau>0$, the NT-Xent term is :
\begin{equation} \label{for:nt-xent} \mathcal{L}_{\text{NT-Xent}}(\mathbf{v}_i, \mathbf{v}_j, \tau) = -\log \left( \frac{\exp(\text{sim}(\mathbf{v}_i, \mathbf{v}_j) / \tau)}{\sum_{k \neq i} \exp(\text{sim}(\mathbf{v}_i, \mathbf{v}_k) / \tau)} \right). \end{equation}
Here, $\text{sim}(\mathbf{v}_i, \mathbf{v}_j)$ denotes the cosine similarity between the projection vectors:
\begin{equation} \text{sim}(\mathbf{v}_i, \mathbf{v}_j) = \frac{\mathbf{v}_i \cdot \mathbf{v}_j}{|\mathbf{v}_i| |\mathbf{v}_j|}. \end{equation}
For each mini-batch containing $N$ anchors, we form two augmented views per anchor and each view is encoded by the shared backbone and then mapped by two independent projection heads tailored to the feature types: we compute the embedding $\mathbf{z}^{(t)}_i=\mathcal{E}(\breve{\mathbf{x}}^{(t)}_i)$, feed it to a continuous-focused projector $\mathcal{P}_{\text{cont}}$ to obtain $\mathbf{v}^{(t)}_{\text{cont},i}=\mathcal{P}_{\text{cont}}(\mathbf{z}^{(t)}_i)$, and to a categorical-focused projector $\mathcal{P}_{\text{cat}}$ to obtain $\mathbf{v}^{(t)}_{\text{cat},i}=\mathcal{P}_{\text{cat}}(\mathbf{z}^{(t)}_i)$ for $t\!\in\!\{1,2\}$ and $i\!=\!1,\dots,N$. Each anchor, therefore, appears twice (once per view) as the “query,” and for each query we use its counterpart view as the positive and all other in-batch projected views from the same head as negatives. To preserve symmetry, we evaluate NT-Xent in both directions (view~1 as anchor with view~2 as positive, and vice versa) and average across the resulting $2N$ anchorings per head.
The head-specific losses are thus
\begin{equation}
\label{eq:ntxent_cont_compact}
\begin{aligned}
\mathcal{L}_{\text{cont}}
&= \frac{1}{2N}\sum_{i=1}^{N}\Big[
\mathcal{L}_{\text{NT-Xent}}\!\big(\mathbf{v}^{(1)}_{\text{cont},i},\mathbf{v}^{(2)}_{\text{cont},i},\tau_{\text{cont}}\big)
\\[-2pt]
&\hspace{32pt}
+\;
\mathcal{L}_{\text{NT-Xent}}\!\big(\mathbf{v}^{(2)}_{\text{cont},i},\mathbf{v}^{(1)}_{\text{cont},i},\tau_{\text{cont}}\big)
\Big].
\end{aligned}
\end{equation}
\begin{equation}
\label{eq:ntxent_cat_compact}
\begin{aligned}
\mathcal{L}_{\text{cat}}
&= \frac{1}{2N}\sum_{i=1}^{N}\Big[
\mathcal{L}_{\text{NT-Xent}}\!\big(\mathbf{v}^{(1)}_{\text{cat},i},\mathbf{v}^{(2)}_{\text{cat},i},\tau_{\text{cat}}\big)
\\[-2pt]
&\hspace{32pt}
+\;
\mathcal{L}_{\text{NT-Xent}}\!\big(\mathbf{v}^{(2)}_{\text{cat},i},\mathbf{v}^{(1)}_{\text{cat},i},\tau_{\text{cat}}\big)
\Big].
\end{aligned}
\end{equation}

The overall contrastive objective combines the two heads with a convex weight:
\begin{equation}
\label{eq:dual_head_final}
\mathcal{L}_{\text{TabCL}}
\;=\;
\lambda\,\mathcal{L}_{\text{cont}}
\;+\;
(1-\lambda)\,\mathcal{L}_{\text{cat}},
\qquad
\lambda\in[0,1].
\end{equation}
In our framework, instead of a single projection head with a single temperature shared across all features, we employ dual-head projections with head-specific temperatures. This design acknowledges that network-flow tables are heterogeneous features.

Using distinct temperatures is beneficial because categorical similarities are typically sharper (best handled with a smaller $\tau_{\text{cat}}$), while continuous similarities are more variable (more stable with a larger $\tau_{\text{cont}}$). The shared encoder thus learns representations that are simultaneously robust for both feature types, while each head specializes its geometry (via temperature) to the statistics of its slice.

\subsubsection{Fine-tuning, pseudo-labeling, and hyperparameters}
After contrastive pretraining, we discard $\mathcal{P}_{\text{cont}}$ and $\mathcal{P}_{\text{cat}}$, attach a classifier $\mathcal{C}(\,\cdot\,;\theta_{\mathcal{C}})$ to $\mathcal{E}$, and fine-tune on $D_s$ using the freeze--unfreeze schedule and cross-entropy (cf. Eq.~\eqref{for:loss_ce}). We then apply $\mathcal{C}\!\circ\!\mathcal{E}$ to all $\mathbf{x}\in D_l$ to obtain final pseudo-labels. The hyperparameters such as
temperatures $(\tau_{\text{cat}},\tau_{\text{cont}})$, the mixing coefficient $\lambda$, the replacement rate $r$, and the refresh schedule are selected on $D_s$ via stratified cross-validation to maximize macro-F1 after fine-tuning. Algorithm \ref{alg:tabcl_dual} details the expanded procedure.

\begin{algorithm}[t]
\caption{TabCL with Class-Conditioned Replacement, Constraint Projection, and Dual-Head Objective}
\label{alg:tabcl_dual}
\begin{algorithmic}[1]
\Require Unlabeled $D_l$, labeled $D_s$; replacement rate $r$; refresh interval; temperatures $\tau_{\text{cont}},\tau_{\text{cat}}$; mixing $\lambda$
\State \textbf{Bootstrap}: train a light classifier on $D_s$; get initial $\{\tilde y_i^{(0)}\}$ for $D_l$
\State Initialize encoder $\mathcal{E}$ and projectors $\mathcal{P}_{\text{cont}},\mathcal{P}_{\text{cat}}$
\For{epoch $=1,2,\dots$}
  \If{refresh interval reached}
    \LongState{Freeze $\mathcal{E}$; train a linear probe on $\{(\mathcal{E}(\mathbf{x}),y)\}_{D_s}$; reclassify $\{\mathcal{E}(\mathbf{x})\}_{D_l}$ to obtain $\{\tilde y_i^{(t)}\}$; stop refreshing if label-change fraction $<$ tolerance or max rounds reached.}
  \EndIf
  \For{mini-batch $\{\mathbf{x}_i\}_{i=1}^{N}\subset D_l$}
    \State For each $i$: set $c_i\gets \tilde y_i^{(t)}$; sample index sets $\mathcal{I}^{(1)},\mathcal{I}^{(2)}$ of size $\lceil rd\rceil$
    \LongState{For $t\in\{1,2\}$: replace $\widetilde{\mathbf{x}}^{(t)}_i[j]\sim \mathbb{P}(\mathbf{x}[j]\mid \tilde y=c_i)$ for $j\in\mathcal{I}^{(t)}$, copy others from $\mathbf{x}_i$; then project to constraints $\breve{\mathbf{x}}^{(t)}_i \gets \Pi_{\mathcal{G}}(\widetilde{\mathbf{x}}^{(t)}_i)$ as in \eqref{eq:tabcl_proj}.}
    \State Encode/project: $\mathbf{z}^{(t)}_i\gets \mathcal{E}(\breve{\mathbf{x}}^{(t)}_i)$;\; $\mathbf{v}^{(t)}_{\text{cont},i}\gets \mathcal{P}_{\text{cont}}(\mathbf{z}^{(t)}_i)$;\; $\mathbf{v}^{(t)}_{\text{cat},i}\gets \mathcal{P}_{\text{cat}}(\mathbf{z}^{(t)}_i)$
  \EndFor
  \State Compute $\mathcal{L}_{\text{cont}}$ and $\mathcal{L}_{\text{cat}}$ via \eqref{eq:ntxent_cont_compact}, \eqref{eq:ntxent_cat_compact}
  \State Compute $\mathcal{L}_{\text{TabCL}}=\lambda\mathcal{L}_{\text{cont}}+(1-\lambda)\mathcal{L}_{\text{cat}}$ \eqref{eq:dual_head_final}; update parameters
\EndFor
\State Discard projectors; attach $\mathcal{C}$ to $\mathcal{E}$; fine-tune on $D_s$ with \eqref{for:loss_ce}; pseudo-label $D_l$
\Ensure Pseudo-labeled $D_l$ from TabCL
\end{algorithmic}
\end{algorithm}

\subsection{Fusion of AE and TabCL Pseudo-Labels}
\label{subsec:fusion}
Either SSL branch (AE or TabCL) can be used on its own to produce a pseudo-labeled pool for the subsequent confident-learning stage. When both are available, we adopt a simple voting rule that consolidates their predictions at negligible overhead. For each flow $\mathbf{x}$, let
\begin{align}
\label{eq:vote_labels}
\widehat{c}^{\text{AE}}(\mathbf{x})
&= \arg\max_{k\in\{1,\dots,K\}}\,p^{\text{AE}}_k(\mathbf{x}), \\[2pt]
\widehat{c}^{\text{TabCL}}(\mathbf{x})
&= \arg\max_{k\in\{1,\dots,K\}}\,p^{\text{TabCL}}_k(\mathbf{x}),
\end{align}
be the top-1 classes predicted by the AE and TabCL classifiers, where $p^{\text{AE}}(\mathbf{x}),p^{\text{TabCL}}(\mathbf{x})\in[0,1]^K$ are their respective softmax probability vectors. Define the \emph{confidence} of each branch as its top-1 probability,
\begin{equation}
\label{eq:vote_conf}
s^{\text{AE}}(\mathbf{x}) \;=\; \max_k\, p^{\text{AE}}_k(\mathbf{x}), 
\qquad
s^{\text{TabCL}}(\mathbf{x}) \;=\; \max_k\, p^{\text{TabCL}}_k(\mathbf{x}),
\end{equation}
and the \emph{margin} (gap between the highest and the second-highest probabilities),
\begin{align}
\label{eq:vote_margin}
m^{\text{AE}}(\mathbf{x}) \;=&\; s^{\text{AE}}(\mathbf{x})
\;-\; \max_{k\neq \widehat{c}^{\text{AE}}} p^{\text{AE}}_k(\mathbf{x}), \\[2pt]
m^{\text{TabCL}}(\mathbf{x}) \;=&\; s^{\text{TabCL}}(\mathbf{x})
\;-\; \max_{k\neq \widehat{c}^{\text{TabCL}}} p^{\text{TabCL}}_k(\mathbf{x}).
\end{align}
The fused pseudo-label $\widehat{c}(\mathbf{x})$ is decided by agreement, then by confidence, then by margin:
\begin{equation}
\label{eq:vote_rule}
\widehat{c}(\mathbf{x}) \;=\;
\begin{cases}
\widehat{c}^{\text{AE}}(\mathbf{x}) & \text{if }\widehat{c}^{\text{AE}}(\mathbf{x})=\widehat{c}^{\text{TabCL}}(\mathbf{x}),\\[4pt]
\widehat{c}^{\text{AE}}(\mathbf{x}) & \text{if } s^{\text{AE}}(\mathbf{x}) > s^{\text{TabCL}}(\mathbf{x}),\\[4pt]
\widehat{c}^{\text{TabCL}}(\mathbf{x}) & \text{if } s^{\text{TabCL}}(\mathbf{x}) > s^{\text{AE}}(\mathbf{x}),\\[4pt]
\widehat{c}^{\text{AE}}(\mathbf{x}) &
\begin{array}{l}
\text{if } s^{\text{AE}}(\mathbf{x}) = s^{\text{TabCL}}(\mathbf{x})\\
\text{and } m^{\text{AE}}(\mathbf{x}) \ge m^{\text{TabCL}}(\mathbf{x})
\end{array}
\\[4pt]
\widehat{c}^{\text{TabCL}}(\mathbf{x}) & \text{otherwise.}
\end{cases}
\end{equation}
This rule favors the model with the higher top-1 confidence; if confidences tie exactly, the larger margin (sharper decision) prevails. In practice, exact ties are exceedingly rare; the last line provides a deterministic fallback. The resulting single pseudo-labeled set is then passed to the downstream task. The process of fusion is detailed in Algorithm \ref{alg:fusion_simple}

\begin{algorithm}[t]
\caption{Confidence--Margin Voting Fusion of AE and TabCL}
\label{alg:fusion_simple}
\begin{algorithmic}[1]
\Require Trained AE and TabCL classifiers; unlabeled $D_l$
\Ensure Fused pseudo-labels on $D_l$
\For{each $\mathbf{x}\in D_l$}
  \State $\mathbf{p}^{\text{AE}}\gets$ softmax from AE;\quad $\mathbf{p}^{\text{TabCL}}\gets$ softmax from TabCL
  \State $\widehat{c}^{\text{AE}}\gets\arg\max_k \mathbf{p}^{\text{AE}}[k]$;\quad $\widehat{c}^{\text{TabCL}}\gets\arg\max_k \mathbf{p}^{\text{TabCL}}[k]$ \eqref{eq:vote_labels}
  \If{$\widehat{c}^{\text{AE}}=\widehat{c}^{\text{TabCL}}$}
     \State $\widehat{c}\gets \widehat{c}^{\text{AE}}$
  \Else
     \State $s^{\text{AE}}\gets \max_k \mathbf{p}^{\text{AE}}[k]$;\quad $s^{\text{TabCL}}\gets \max_k \mathbf{p}^{\text{TabCL}}[k]$ \eqref{eq:vote_conf}
     \State $m^{\text{AE}}\gets s^{\text{AE}}-\max_{k\neq \widehat{c}^{\text{AE}}}\mathbf{p}^{\text{AE}}[k]$;\quad
           $m^{\text{TabCL}}\gets s^{\text{TabCL}}-\max_{k\neq \widehat{c}^{\text{TabCL}}}\mathbf{p}^{\text{TabCL}}[k]$ \eqref{eq:vote_margin}
     \LongState{Decide $\widehat{c}$ by rule \eqref{eq:vote_rule}: prefer higher confidence; if tied, prefer larger margin; fall back deterministically otherwise.}
  \EndIf
  \State Assign $\widehat{c}$ to $\mathbf{x}$
\EndFor
\end{algorithmic}
\end{algorithm}

Fusion often improves corner cases where the two branches disagree on difficult classes, with minimal changes to the pipeline. The cost is additional SSL pretraining (both branches), which we report explicitly in the efficiency section, together with parameter counts and epochs for each path.

\subsection{Confident Learning for Noise Reduction and Final Classification}
The pseudo-labeled pool $D_l=\{(\mathbf{x}_i,\tilde y_i)\}_{i=1}^{|D_l|}$ produced by SSL, as detailed in the preceding subsections, inevitably contains noise. Training directly on these labels degrades performance. We aim to reduce the impact of noisy pseudo-labels before training the final classifier. We therefore adopt and extend Confident Learning (CL)~\cite{26} into a pipeline that down-weights noisy pseudo-labels and trains the final classifier in five stages: First, following standard CL practice, we compute out-of-sample prediction probabilities and define each sample’s self-confidence. Second, we introduce class-adaptive, quantile-based confidence thresholds with robust dispersion (MAD) to reflect per-class difficulty. Third, we convert confidence into smooth, calibration-aware per-sample logistic weights that attenuate uncertain pseudo-labels while retaining all data. Fourth, leveraging the CL confident joint, we perform a class-wise balanced retention calibration that aligns each class’s retained mass with its estimated clean fraction, mitigating imbalance under label noise. Finally, we train the final classifier $\mathcal{C}_{\text{final}}$ using weighted symmetric cross-entropy on the full pseudo-labeled set. Importantly, we do not remove samples; rather, we reduce the contribution of uncertain ones and calibrate class masses to stabilize learning. The complete CL-driven noise mitigation and downstream training procedure is summarized in Algorithm~\ref{alg:cl_weighted}.

\begin{algorithm}[t]
\caption{Traffic-Adapted CL with Quantile Thresholds, Logistic Weights, BRC, and Weighted SCE}
\label{alg:cl_weighted}
\begin{algorithmic}[1]
\Require Pseudo-labeled $D_l=\{(\mathbf{x}_i,\tilde y_i)\}$; small labeled $D_s$; folds $F$; grids $\mathcal{Q}$ (quantile $q$), $\mathcal{W}$ ($w_{\min}$), $\Gamma$ (slope $\gamma$)
\Ensure Trained final classifier $\mathcal{C}_{\text{final}}$
\State \textbf{Out-of-sample probabilities}: obtain $\mathbf{p}_i$ via $F$-fold CV \eqref{eq:probvec}; set $s_i\gets \mathbf{p}_i[\tilde y_i]$ \eqref{eq:selfconf}
\State \textbf{Thresholds/weights per class}:
\For{$q\in\mathcal{Q}$,\ $w_{\min}\in\mathcal{W}$,\ $\gamma\in\Gamma$}
   \For{each class $j$}
     \State Compute $t^{(q)}_j$ \eqref{eq:quantile}, $\tilde m_j$ \eqref{eq:median}, $\mathrm{MAD}_j$ \eqref{eq:mad}, $\sigma_j$ \eqref{eq:sigma_clip}
   \EndFor
   \For{each $i$}
     \State $z_i \gets (s_i - t^{(q)}_{\tilde y_i}) / (\gamma\,\sigma_{\tilde y_i})$ \eqref{eq:zscore}
     \State $w_i \gets w_{\min} + (1-w_{\min})\cdot \mathrm{sigm}(z_i)$ \eqref{eq:weight}
   \EndFor
   \State Build confident joint $\widehat{Q}$ \eqref{eq:Qhat}; get $\rho_j$ \eqref{eq:rho}; compute $a_j$ \eqref{eq:aj}
   \State Final weights: $w_i' \gets \operatorname{clip}(a_j\,w_i,\,w_{\min},\,1)$ \eqref{eq:final_weight} for $\tilde y_i=j$
   \State Train $\mathcal{C}_{\text{final}}$ on $D_l$ with weighted SCE \eqref{eq:SCE_per_sample}
\EndFor
\LongState{\textbf{Deployment:} fix $(q,w_{\min},\gamma)$ according to the chosen configuration; recompute $\{w_i'\}$ on the full pseudo-labeled pool and train $\mathcal{C}_{\text{final}}$ on $D_l$.}
\end{algorithmic}
\end{algorithm}

\subsubsection{Prediction Probabilities and Self-Confidence}
We first obtain out-of-sample predicted probabilities $\mathbf{p}_i$ for each sample $\mathbf{x}_i$ in $D_l$ using $F$-fold cross-validation with a simple classifier, for each sample $\mathbf{x}_i$, use the model trained on the $F\!-\!1$ folds that did not include $\mathbf{x}_i$. Denoting the parameters of this model by $\theta^{(-f(i))}$, we calculate the $K$-way probability vector:

\begin{equation}
\label{eq:probvec}
\begin{aligned}
\mathbf{p}_i \;=\; \big[\, & P\!\big(y{=}1 \mid \mathbf{x}_i;\theta^{(-f(i))}\big),\, \ldots,
P\!\big(y{=}K \mid \mathbf{x}_i;\theta^{(-f(i))}\big) \,\big], \\
& \mathbf{p}_i \in [0,1]^K, \qquad \sum_{k=1}^{K}\mathbf{p}_i[k]=1 .
\end{aligned}
\end{equation}

In other words, $\mathbf{p}_i$ is $P(y\mid \mathbf{x}_i)$ evaluated for all classes. The top-1 class of this out-of-sample model is $\hat y_i^{\star}=\arg\max_{k}\mathbf{p}_i[k]$, which may or may not equal the SSL pseudo-label $\tilde y_i$. In CL, the quantity of interest is self-confidence $s_i$, defined as the probability that the out-of-sample model assigns to the observed (pseudo) label of the sample:

\begin{equation}
\label{eq:selfconf}
s_i \;=\; \mathbf{p}_i[\tilde y_i] \;\in\; [0,1].
\end{equation}

Thus, $\mathbf{p}_i$ is a $K$-dimensional probability vector ($P(y\mid \mathbf{x}_i)$ across all classes), while $s_i$ is a single scalar entry of that vector corresponding to $\tilde y_i$. These out-of-sample predictions and the resulting self-confidences are used exclusively to vet pseudo-labels via CL; they do not provide gradients to $\mathcal{C}_{\text{final}}$.

\subsubsection{Per-class, quantile-calibrated thresholds}
For each observed class $j\in\{1,\dots,K\}$, collect the self-confidence scores of samples whose pseudo-label equals $j$: $\mathcal{S}_j=\{\,s_i: \tilde y_i=j\,\}$, and $n_j=|\mathcal{S}_j|$. Let the elements of $\mathcal{S}_j$ in non-decreasing order be
$s_{j,(1)} \le s_{j,(2)} \le \cdots \le s_{j,(n_j)}$ (order statistics).
For a quantile level $q\in(0,1]$, define the per-class $q$-quantile threshold by the standard order-statistic rule:
\begin{equation}
\label{eq:quantile}
\begin{aligned}
t^{(q)}_j \;&=\; \mathrm{Quantile}_q(\mathcal{S}_j) \;\triangleq\; s_{j,\,(k_j(q))},\\
k_j(q) \;&=\; \left\lceil q\,n_j \right\rceil, \qquad q\in(0,1] .
\end{aligned}
\end{equation}
(Equivalently, $t^{(q)}_j$ is the smallest value for which at least a fraction $q$ of the scores in $\mathcal{S}_j$ are $\le t^{(q)}_j$; linear interpolation can be used if a specific quantile convention is desired.)

We treat $q$ as a hyperparameter shared across classes and select it by stratified cross-validation on the small labeled set $D_s$. This procedure ties the threshold choice directly to downstream performance while preserving the per-class adaptivity of $t^{(q)}_j$.

Canonical CL~\cite{26} uses the per-class mean self-confidence as a threshold $t^{(\mathrm{mean})}_j \;=\; \bar{S}_j $. In contrast, we set the class threshold to a quantile of $\mathcal{S}_j$, which offers two advantages for network traffic flows: (i) it adapts to class-dependent score distributions that are often skewed due to heterogeneous or encrypted traffic patterns, making the cutoff robust to outliers; and (ii) it provides an explicit strictness control via $q$: higher $q$ yields a stricter threshold (fewer samples lie above $t^{(q)}_j$ and will later receive large weights), while lower $q$ is more permissive.

\subsubsection{Calibration-aware per-sample weights}
To avoid hard pruning (outright removal of suspected-noisy samples) and to stabilize training, we define a smooth weight for each sample that depends on its distance to the class threshold and on the dispersion of confidences in that class. For each observed class, we define the class median by:
\begin{equation}
\label{eq:median}
\tilde m_j \;=\; \mathrm{median}\big(\mathcal{S}_j\big).
\end{equation}
We measure dispersion using the \emph{Median Absolute Deviation (MAD)}:
\begin{equation}
\label{eq:mad}
\mathrm{MAD}_j \;=\; \mathrm{median}\big\{\,|s-\tilde m_j| : s\in\mathcal{S}_j \,\big\}.
\end{equation}
We set the classwise scale to $\sigma_j$ and clip it to avoid division by zero:
\begin{equation}
\label{eq:sigma_clip}
\sigma_j \;\leftarrow\; \max( \mathrm{MAD}_j,\varepsilon), \quad \varepsilon>0.
\end{equation}
Using MAD makes the scale estimate robust to outliers and skewed score distributions common in traffic flows. Given the per-class quantile threshold $t^{(q)}_j$ from \eqref{eq:quantile}, define for sample $i$ with $\tilde y_i=j$:
\begin{equation}
\label{eq:zscore}
z_i \;=\; \frac{s_i - t^{(q)}_{\tilde y_i}}{\gamma\, \sigma_{\tilde y_i}},
\end{equation} where $\gamma>0$ controls the slope around the threshold. We map $z_i$ to $(0,1)$ with the logistic function that is:
\begin{equation}
\label{eq:sigmoid}
\operatorname{sigm}(z_i) \;=\; \frac{1}{1 + e^{-z_i}}.
\end{equation} With a minimum weight $w_{\min}\in(0,1)$, the per-sample training weight is: 
\begin{equation}
\label{eq:weight}
w_i \;=\; w_{\min} \;+\; (1 - w_{\min}) \cdot \operatorname{sigm}(z_i)
\;\;\in\; [\,w_{\min},\,1\,].
\end{equation} All hyperparameters in \eqref{eq:sigma_clip}–\eqref{eq:weight} (e.g., $w_{\min}$, $\gamma$, $\varepsilon$) are selected via stratified cross-validation on the small labeled set $D_s$. 

 The weighting mechanism in \eqref{eq:weight} behaves in a stable and interpretable manner. By construction $w_i\in[w_{\min},1]$; When $s_i \!\ll\! t^{(q)}_{\tilde y_i}$, the pseudo-label of the flow is likely noisy, the standardized margin $z_i\!\to\!-\infty$ and thus $w_i\!\to\!w_{\min}$. Conversely, when $s_i \!\gg\! t^{(q)}_{\tilde y_i}$, the pseudo-label is likely correct, $z_i\!\to\!+\infty$ and $w_i\!\to\!1$. At the threshold, if $s_i=t^{(q)}_{\tilde y_i}$ then $z_i=0$ and $w_i=w_{\min}+\tfrac{1}{2}(1-w_{\min})$, i.e., the exact midpoint between the minimum and maximum weights.

 \subsubsection{Confident joint and class-wise balanced retention}
 In CL, the matrix $\widehat{Q}\in\mathbb{R}^{K\times K}$ summarizes the estimated joint between the latent true class (rows) and the pseudo-label (columns). It is computed directly from out-of-sample probability vectors $\mathbf{p}_i$ using soft counts. Specifically,
\begin{equation}
\label{eq:Qhat}
\widehat{Q}_{y=k,\;\tilde y=j} \;=\; \sum_{i:\,\tilde y_i=j} \mathbf{p}_i[k], \qquad k,j\in\{1,\dots,K\},
\end{equation}
where the row index “$y=k$’’ denotes the latent class inferred from $\mathbf{p}_i$. The diagonal entry $\widehat{Q}_{j,j}$ estimates the clean mass for class $j$, while the column sum $\sum_k \widehat{Q}_{k,j}$ is the total mass observed as $j$. The corresponding estimated clean fraction for observed class $j$ is:
\begin{equation}
\label{eq:rho}
\rho_j \;=\; \frac{\widehat{Q}_{j,j}}{\sum_{k}\widehat{Q}_{k,j}} \;\in\; [0,1].
\end{equation}
To mitigate minority erosion under class imbalance, a Balanced Retention Constraint (BRC) is applied after computing per-sample weights $w_i$. Define $n_j \triangleq \#\{i:\tilde y_i=j\}$, the target effective mass $T_j \triangleq \rho_j\,N_j$, and the current mass $M_j \triangleq \sum_{i:\,\tilde y_i=j} w_i$. The classwise scaling factor is:
\begin{equation}
\label{eq:aj}
a_j \;=\; \frac{T_j}{\max(\epsilon,\,M_j)} ,
\end{equation}
where $\epsilon>0$ provides numerical stability. The final per-sample weight used in training is then:
\begin{equation}
\label{eq:final_weight}
w_i' \;=\; \operatorname{clip}\!\big(a_j\,w_i,\,w_{\min},\,1\big),
\end{equation}
This soft constraint yields $\sum_{i:\,\tilde y_i=j} w_i' \approx T_j$ (within a small tolerance), aligning the retained mass of each observed class with its clean fraction estimated from \eqref{eq:Qhat}–\eqref{eq:rho}.

If $M_j<T_j$, then $a_j>1$ and all weights in class $j$ are uniformly lifted (subject to clipping), preventing under-retention of harder/minority classes; if $M_j>T_j$, then $a_j<1$ and the class is gently down-scaled to avoid over-retention. Because $w_i\in[w_{\min},1]$, the effective mass per class lies in $[w_{\min}n_j,\,n_j]$; the BRC moves it toward $T_j=\rho_j N_j$, thereby preserving minority classes in proportion to their estimated cleanliness. Equation~\eqref{eq:Qhat} follows the standard confident-joint construction in CL~\cite{26}, whereas the scaling rule in \eqref{eq:aj} and  \eqref{eq:final_weight} constitutes the proposed extension that operationalizes class-aware retention during training.

\subsubsection{Final training with weighted SCE}
The downstream classifier $\mathcal{C}_{\text{final}}$ is trained on the pseudo-labeled dataset $D_l=\{(\mathbf{x}_i,\tilde y_i)\}$ using the Symmetric Cross-Entropy (SCE) loss, weighted by the final per-sample weight $w_i'$ from \eqref{eq:final_weight}. For a sample $(\mathbf{x}_i,\tilde y_i)$ with predicted probability vector $\mathbf{p}_i=\mathcal{C}_{\text{final}}(\mathbf{x}_i)\in[0,1]^K$, the loss is:
\begin{equation}
\label{eq:SCE_per_sample}
\mathcal{L}_{\text{SCE}}^{(i)} \;\triangleq\; w_i'\,\big[\,\alpha\,\mathrm{CE}(\mathbf{p}_i,\tilde y_i)\;+\;\beta\,\mathrm{RCE}(\tilde y_i,\mathbf{p}_i)\,\big].
\end{equation}
The coefficients $\alpha,\beta>0$ are hyperparameters and are selected by stratified cross-validation on the small labeled set $D_s$.
We adopt SCE because it combines the standard cross-entropy (encouraging correct predictions) with a reverse cross-entropy term that penalizes overconfident mistakes, improving robustness under residual label noise in pseudo-labeled data. Concretely,
\begin{equation}
\label{eq:label_smooth}
\hat y_{i,k} \;=\;
\begin{cases}
1-\epsilon, & k=\tilde y_i,\\[2pt]
\dfrac{\epsilon}{K-1}, & k\neq \tilde y_i,
\end{cases}
\qquad \epsilon \in (0,1).
\end{equation}

\begin{equation}
\label{eq:SCE_terms_correct}
\begin{aligned}
\mathrm{CE}(\mathbf{p}_i,\tilde y_i) \;&=\; -\sum_{k=1}^K \hat y_{i,k}\,\log \mathbf{p}_i[k],\\
\mathrm{RCE}(\tilde y_i,\mathbf{p}_i) \;&=\; -\sum_{k=1}^K \mathbf{p}_i[k]\;\log \hat y_{i,k}.
\end{aligned}
\end{equation}

Then, parameters of $\mathcal{C}_{\text{final}}$ are updated by gradient descent:
\begin{equation}
\label{for:update}
\theta_{\text{final}} \;\leftarrow\; \theta_{\text{final}} \;-\; \eta\,\nabla_{\theta_{\text{final}}}\,\mathcal{L}_{\text{SCE}}.
\end{equation}

\section{Evaluation \& Results}
\label{sec:results} 
The evaluation of the proposed network traffic classification model is presented in this section, with the objective of assessing its performance. The aim is to compare the model's effectiveness using key metrics such as precision, F1-score, recall, and classification accuracy.

\subsection{Simulation Setting and Datasets}
\label{subsec:A}
To assess the proposed model, we utilized two distinct datasets: a self-generated dataset treated as unlabeled for self-supervised learning and a small labeled dataset derived from the ISCX VPN-nonVPN dataset \cite{12}. We used a self-generated dataset to demonstrate the full, deployment-style pipeline from raw packet capture to flow-level tabular features under realistic label scarcity and class imbalance. In addition to these datasets, we also evaluate on the UCDavis-QUIC \cite{44} dataset to broaden the evaluation and assess generalization to modern encrypted traffic. Below, we describe each dataset in detail, including how it was created, its characteristics, and its role in the evaluation.

\begin{figure}[h]
    \centering
    \includegraphics[width=\columnwidth]{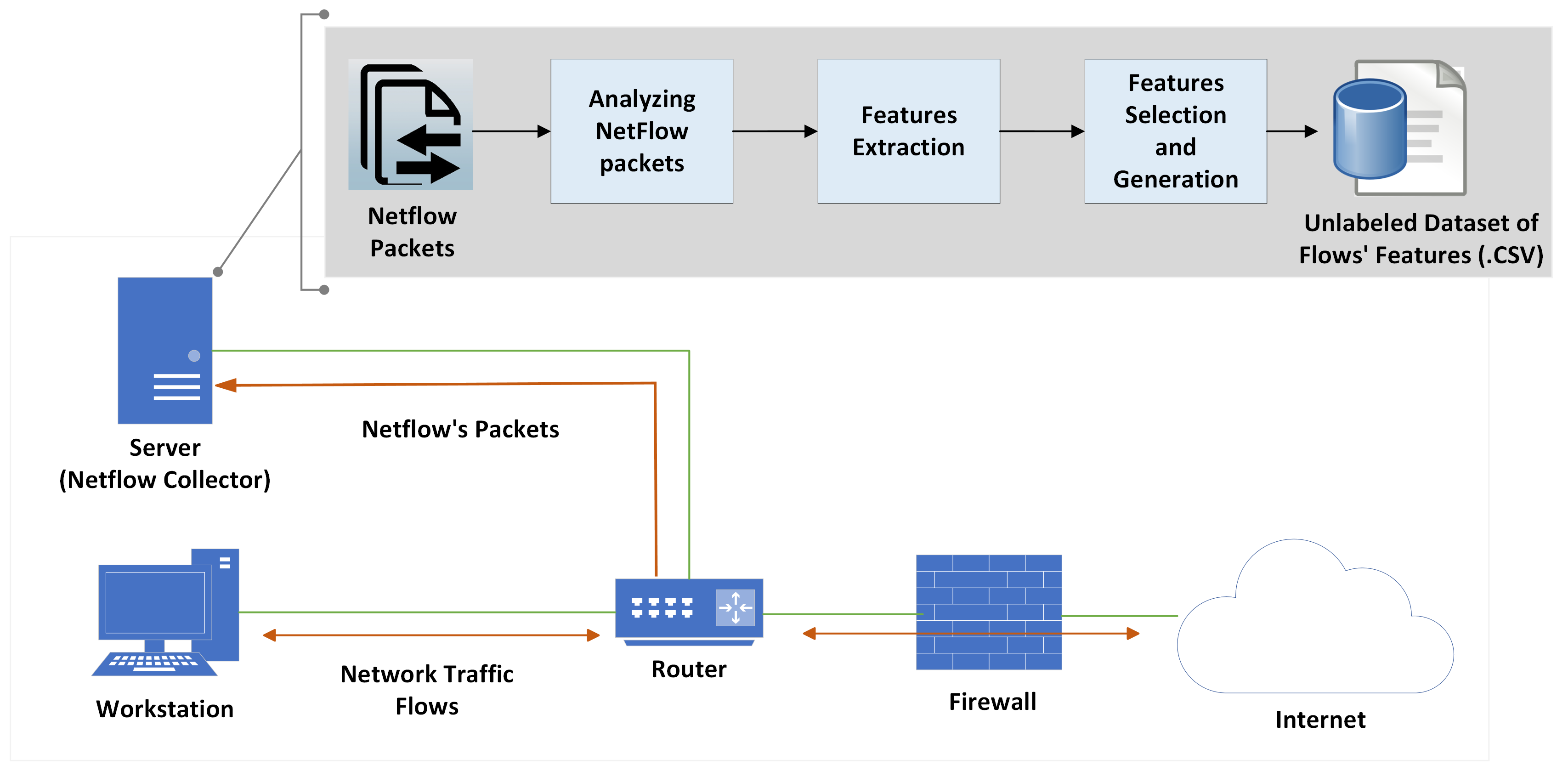}
    \caption{Diagram of the network architecture and Flowchart of the unlabeled dataset generation using NetFlow.}
    \label{fig:Netflow_Architecture}
\end{figure}

\textbf{Self-Generated Dataset as an Unlabeled:} The large unlabeled dataset was created using GNS3 (Global Network Simulator 3), a widely used tool for simulating network environments. In this setup, we configured a simulated Cisco router with NetFlow, a service that collects and monitors IP traffic data as it passes through the router. The router acted as a NetFlow exporter, sending traffic information to a virtual server set up as a NetFlow collector. This virtual server, operating within a VMware environment, processed NetFlow packets sent from the router, extracting features from the traffic generated by a Windows 10 virtual machine workstation running 10 distinct applications. These applications were carefully selected to represent different types of network traffic: VPN traffic (e.g., VPN\_Vimeo, Tor\_YouTube), time-sensitive traffic (e.g., Skype\_VideoCall, Facebook\_Audio), file transfer traffic (e.g., FTPS\_Upload, SFTP\_Upload, SCP\_Download), and video/voice traffic (e.g., YouTube, Skype\_Chat, Email). This variety ensures the dataset covers a broad range of network behaviors seen in real-world scenarios. The combined network architecture and dataset-generation workflow are shown in Fig.~\ref{fig:Netflow_Architecture}. 

The captured NetFlow data were then processed using a custom Python script that follows this workflow on the NetFlow collector, handling both feature extraction and feature generation. The script extracted statistical features, like packet count, octet count, and throughput, along with time-based attributes such as duration and inter-arrival times. In the "generation" step, the script also processed raw NetFlow packets to calculate additional features (e.g., minimum, maximum, mean, and standard deviation of some features). The final dataset, saved as a Comma-Separated Values (CSV) file, contains 21 features and 8,684 flows. The combined corpus is intentionally imbalanced, reflecting real traffic mixes: YouTube accounts for 41.9\% of flows, while minority classes such as SFTP\_Upload and SCP\_Download are each under 4.5\%. This skew is typical in NTC and is handled in our pipeline by class-aware thresholds and balanced retention in the CL stage. Feature distributions are physically plausible and heavy-tailed, as expected for flows: the median Mean Packet Length is 571 B (95th percentile = 1,406 B, below the Ethernet MTU), the median Mean IAT is 6.9 ms (95th percentile = 0.59 s).

\textbf{ISCX VPN-nonVPN as a Small Labeled Dataset:} The smaller labeled dataset was derived from the ISCX VPN-nonVPN dataset \cite{12}, a well-known public dataset developed for network traffic classification research. This dataset is originally provided in packet-level format as .pcap files, which contain detailed captures of network packets from various activities. It includes a mix of VPN and non-VPN traffic, covering both encrypted and unencrypted flows across different applications and conditions. This diversity makes it a strong benchmark for testing traffic classification models in realistic scenarios, especially where encryption and varying network states are involved. To adapt this dataset for our needs, we used custom Python scripts to process the .pcap files and extract flow-level traffic data. This involved analyzing packet sequences, grouping them into flows based on attributes like source/destination IP addresses and port numbers, and selecting a subset of 538 flows, chosen to limit labeling effort and ensure only complete flows were included. These flows match the same 10 application categories as the large dataset, ensuring consistency. This smaller dataset serves as a labeled resource for fine-tuning the model, despite its very smaller size compared to the self-generated dataset. To reflect real-world traffic distributions, the number of flows per class was intentionally varied, resulting in an imbalanced dataset. This distribution is detailed in Table \ref{tab:flow_counts}, which enumerates the flow counts for each class.

\begin{table}[h]
    \centering
    \caption{Number of flows per class in the self-generated dataset and the ISCX VPN-nonVPN dataset, with class share over the combined corpus.}
    \label{tab:flow_counts}
    \begin{tabular}{lccc}
        \toprule
        \multirow{2}{*}{\textbf{Class}} & \multicolumn{2}{c}{\textbf{Number of Flows}} & \multirow{2}{*}{\textbf{Share (\%)}} \\
        \cline{2-3}
        & \text{self gen. Dataset} & \text{ISCX Dataset} & \\
        \midrule
        YouTube          & 3764 & 84 & 41.85 \\
        Skype\_VideoCall & 438  & 63 & 5.44  \\
        Skype\_Chat      & 956  & 41 & 10.80 \\
        FTPS\_Upload     & 638  & 53 & 7.48  \\
        SFTP\_Upload     & 321  & 20 & 3.69  \\
        SCP\_Download    & 387  & 22 & 4.44  \\
        Facebook\_Audio  & 497  & 59 & 6.04  \\
        Email            & 904  & 98 & 10.86 \\
        Tor\_YouTube     & 359  & 44 & 4.38  \\
        VPN\_Vimeo       & 416  & 53 & 5.01  \\
        \midrule
        \textbf{Total}   & \textbf{8684} & \textbf{538} & \textbf{100.00} \\
        \bottomrule
    \end{tabular}
\end{table}

\textbf{UCDavis-QUIC Dataset:} We additionally evaluate on the UCDavis-QUIC dataset, which contains QUIC-only traffic from five Google services (Google Docs, Google Drive, Google Music, Google Search, and YouTube). The released data are packet captures (PCAP) from which flows are derived; in our pipeline we convert PCAP to CSV and aggregate by 5-tuple to obtain per-flow statistics. Consistent with the original paper of the dataset \cite{44}, we employ a flow-level feature schema of 24 numerical statistics per flow. In addition, we retain an explicit categorical flow-direction indicator (forward/backward/bidirectional) alongside the numeric features. The QUIC dataset contains 6,439 flows in total. To mirror our label-scarce setting, we treat only 5\% of flows per class as labeled and the remaining 95\% as unlabeled, like previous datasets. Table \ref{tab:quic_class_dist} reports the class distribution and the stratified labeled/unlabeled split, making the dataset’s imbalance explicit.

\begin{table}[h]
    \centering
    \caption{UCDavis--QUIC class distribution.}
    \label{tab:quic_class_dist}
    \begin{tabular}{lcccc}
        \toprule
        \textbf{Class} & \textbf{Total Flows} & \textbf{Labeled} & \textbf{Unlabeled} & \textbf{Share (\%)} \\
        \midrule
        Google Docs   & 1221 & 61 & 1160 & 18.96 \\
        Google Drive  & 1634 & 82 & 1552 & 25.38 \\
        Google Music  &  592 & 30 &  562 &  9.19 \\
        Google Search & 1915 & 96 & 1819 & 29.74 \\
        YouTube       & 1077 & 54 & 1023 & 16.73 \\
        \midrule
        \textbf{Total} & \textbf{6439} & \textbf{323} & \textbf{6116} & \textbf{100.00} \\
        \bottomrule
    \end{tabular}
\end{table}

\begin{table*}[t]
\centering
\caption{Representative grid search results for AE's hyperparameters.}
\label{tab:gs_ae}
\resizebox{\textwidth}{!}{%
\begin{tabular}{cccccccccc}
\toprule
\textbf{Latent dim} & \textbf{Depth} & \textbf{Hidden width per layer} & \textbf{Dropout} & \textbf{LR} & \textbf{Weight decay} & \textbf{Batch} & \textbf{Epochs} & \textbf{$\boldsymbol{\phi}$ (constraint)} & \textbf{Macro-F1 (\%)} \\
\midrule
128  & 3 & (256, 128, 64) & 0.10 & $5\times 10^{-4}$ & $1\times 10^{-5}$ & 128 & 100 & 0.5 & $\mathbf{93.8\ \pm\ 0.6}$ \\
64 & 2 & (256, 128)     & 0.20 & $5\times 10^{-4}$ & $1\times 10^{-5}$ & 256 & 100 & 0.1 & $92.6\ \pm\ 0.6$ \\
32  & 3 & (256, 128, 64) & 0.00 & $1\times 10^{-3}$ & $1\times 10^{-6}$ & 128 & 50 & 1.0 & $91.2\ \pm\ 0.9$ \\
64  & 2 & (128, 64)      & 0.30 & $1\times 10^{-3}$ & $1\times 10^{-4}$ &  64 & 50 & 0.0 & $89.7\ \pm\ 0.9$ \\
128 & 3 & (256, 128, 64) & 0.10 & $2\times 10^{-4}$ & $1\times 10^{-5}$ & 256 & 80 & 1.0 & $92.9\ \pm\ 0.5$ \\
64  & 3 & (256, 128, 64) & 0.20 & $5\times 10^{-4}$ & $1\times 10^{-4}$ &  64 & 50 & 2.0 & $90.1\ \pm\ 0.8$ \\
\bottomrule
\end{tabular}%
}
\end{table*}

\begin{table*}[t]
\centering
\caption{Representative grid search results for TabCL's hyperparameters.}
\label{tab:gs_tabcl}
\resizebox{\textwidth}{!}{%
\begin{tabular}{ccccccccccccc}
\toprule
\textbf{Depth} & \textbf{Hidden widths} & \textbf{LR} & \textbf{Weight decay} & \textbf{Batch} & \textbf{Epochs} & $\boldsymbol{\tau_{\mathrm{cont}}}$ & $\boldsymbol{\tau_{\mathrm{cat}}}$ & $\boldsymbol{r}$ & $\boldsymbol{\lambda}$ & \textbf{Proj dim} & \textbf{Refresh} & \textbf{Macro-F1 (\%)} \\
\midrule
3 & (256, 128, 64) & $5\times 10^{-4}$ & $1\times 10^{-5}$ & 256 & 100 & 0.5 & 0.20 & 0.15 & 0.5 & 128 & 10  & $\mathbf{94.2\ \pm\ 0.5}$ \\
2 & (256, 128)     & $1\times 10^{-3}$ & $1\times 10^{-5}$ & 128 & 80 & 0.7 & 0.20 & 0.10 & 0.5 & 128 & 5 & $92.7\ \pm\ 0.6$ \\
3 & (256, 128, 64) & $1\times 10^{-4}$ & $1\times 10^{-6}$ & 256 & 100 & 0.5 & 0.10 & 0.10 & 0.7 & 256 & 5  & $94.0\ \pm\ 0.6$ \\
2 & (128, 64)      & $5\times 10^{-4}$ & $1\times 10^{-4}$ & 512 &  50 & 0.5 & 0.10 & 0.05 & 0.4 &  64 & 5  & $91.8\ \pm\ 0.8$ \\
3 & (256, 128, 64) & $5\times 10^{-4}$ & $1\times 10^{-5}$ & 128 & 100 & 0.3 & 0.05 & 0.10 & 0.5 & 128 & 10 & $93.6\ \pm\ 0.7$ \\
3 & (256, 128, 64) & $5\times 10^{-4}$ & $1\times 10^{-5}$ & 256 & 100 & 0.7 & 0.20 & 0.20 & 0.5 & 128 & 5  & $88.3\ \pm\ 1.3$ \\
\bottomrule
\end{tabular}%
}
\end{table*}

\begin{table*}[t]
\centering
\caption{Representative grid search results for hyperparameters of Confident Learning (CL) and Final Classifier.}
\label{tab:gs_cl}
\resizebox{\textwidth}{!}{%
\begin{tabular}{cccccccccccccc}
\toprule
\multicolumn{4}{c}{\textbf{CL hyperparameters}} & \multicolumn{9}{c}{\textbf{Final classifier (MLP, weighted SCE)}} & \multirow{2}{*}{\textbf{Macro-F1 (\%)}} \\
\cmidrule(lr){1-4}\cmidrule(lr){5-13}
$\boldsymbol{F}$-Folds & $\boldsymbol{q}$ & $\boldsymbol{w_{\min}}$ & $\boldsymbol{\gamma}$ & \textbf{Depth} & \textbf{Hidden widths} & \textbf{LR} & \textbf{Weight decay} & \textbf{Batch} & \textbf{Dropout} & \textbf{Epochs} & $\boldsymbol{\alpha}$ & $\boldsymbol{\beta}$ & \\
\midrule
5  & 0.70 & 0.20 & 4 & 3 & (512, 256, 128)  & $1\times 10^{-4}$ & $1\times 10^{-5}$ & 128 & 0.30 & 50 & 0.3 & 2.0 & $\mathbf{95.3\ \pm\ 0.5}$ \\
5  & 0.70 & 0.10 & 4 & 3 & (512, 256, 128)  & $5\times 10^{-4}$ & $1\times 10^{-5}$ & 128 & 0.30 & 80 & 0.3 & 2.0 & $95.0\ \pm\ 0.7$ \\
3  & 0.70 & 0.20 & 2 & 2 & (256, 128)      & $5\times 10^{-4}$ & $1\times 10^{-5}$ & 256 & 0.25 & 30 & 0.5 & 1.0 & $94.6\ \pm\ 0.7$ \\
5  & 0.60 & 0.30 & 8 & 3 & (1024, 512, 256) & $3\times 10^{-4}$ & $1\times 10^{-4}$ & 128 & 0.40 & 60 & 0.3 & 3.0 & $94.1\ \pm\ 0.8$ \\
10 & 0.70 & 0.20 & 4 & 3 & (512, 256, 128)  & $1\times 10^{-3}$ & $1\times 10^{-6}$ &  64 & 0.20 & 60 & 0.1 & 2.0 & $93.9\ \pm\ 0.9$ \\
5  & 0.80 & 0.10 & 8 & 2 & (256, 128)       & $1\times 10^{-3}$ & $1\times 10^{-4}$ & 256 & 0.20 & 30 & 0.1 & 1.0 & $93.2\ \pm\ 0.6$ \\
\bottomrule
\end{tabular}%
}
\end{table*}

We conducted our experiments on a workstation running Ubuntu 22.04.4 LTS, Python 3.12.11, and PyTorch 2.8.0+cu126, equipped with an Intel Xeon CPU @ 2.00 GHz, 12.7 GB RAM, and a single NVIDIA Tesla T4 (15.0 GB) GPU.

\subsection{Performance of the Proposed Model on Self-generated and ISCX datasets}
\label{subsec:B}
This subsection evaluates the proposed model's performance on Self-generated and ISCX datasets, focusing on the accuracy of pseudo-label generation via SSL and the final classification accuracy achieved after applying CL to mitigate noisy pseudo-labels.  During training, we treated the self-generated dataset as unlabeled, using its assigned labels only for evaluation purposes. By keeping the dataset unlabeled during training, we relied on SSL to uncover patterns from the plentiful unlabeled data. Meanwhile, the small labeled ISCX VPN-nonVPN dataset was used to fine-tune the model. This combination effectively addresses the scarcity of labeled data while leveraging the strengths of both datasets.

The model's performance is measured using the following metrics. In these metrics, TP, TN, FP, and FN denote true positives, true negatives, false positives, and false negatives, respectively.
\begin{itemize}
    \item \textbf{Accuracy}: The proportion of correctly classified samples. This metric provides an overall measure of the model’s correctness across all classes, defined as:
    
    \begin{equation}
    \text{Accuracy} = \frac{\text{TP} + \text{TN}}{\text{TP} + \text{TN} + \text{FP} + \text{FN}}.
    \end{equation}
    
    \item \textbf{Recall}: The ratio of correctly identified positive samples. Recall indicates the model’s ability to detect all relevant instances of a class, given by: 

    \begin{equation}
    \text{Recall} = \frac{\text{TP}}{\text{TP} + \text{FN}}.
    \end{equation}

    \item \textbf{Precision}: The proportion of correctly predicted positive samples among all samples predicted as positive, calculated as:
    \begin{equation}
    \text{Precision} = \frac{\text{TP}}{\text{TP} + \text{FP}}.
    \end{equation}
    
    \item \textbf{F1-Score}: The harmonic mean of precision and recall. The F1-score balances precision and recall, offering a single metric that reflects the performance of the model, computed as: \begin{equation}
    \text{F1-Score} = 2 \times \frac{\text{Precision} \times \text{Recall}}{\text{Precision} + \text{Recall}}.
    \end{equation}

\end{itemize}
The above metrics collectively offer a robust assessment of the model's classification capabilities. 

About the hyperparameter selection, we tuned each module with a stratified 5-fold cross-validation (CV) on the small labeled ISCX split; for every candidate configuration, we repeated training with 3 random seeds and reported Macro-F1 as mean ± 95\% CI over the resulting 15 scores (5 folds × 3 seeds). We then selected the best settings per module by Macro-F1 and carried those forward to the end-to-end pipeline. Tables \ref{tab:gs_ae}–\ref{tab:gs_cl} present representative configurations from the full grid used to select hyperparameters for AE, TabCL, CL, and the final classifier. Finally, Table \ref{tab:hyperparameters} presents the architecture and hyperparameters for the models utilized in the evaluation section.

\begin{table}[!h]
    \centering
    \caption{Hyperparameters of Models (final choices used in evaluation).}
    \label{tab:hyperparameters}
    \resizebox{\columnwidth}{!}{
    \begin{tabular}{lcccc}
        \toprule
        \textbf{Hyperparameter} & \textbf{Autoencoder} & \textbf{TabCL} & \textbf{Confident Learning} & \textbf{Final Classifier} \\
        \midrule
        \textbf{Model Architecture} & & & & \\
        Hidden Layers & 3 & 3 & 2 & 3 \\
        Neurons in 1\textsuperscript{st} HL & 256 & 256 & 128 & 512 \\
        Neurons in 2\textsuperscript{nd} HL & 128 & 128 & 64 & 256 \\
        Neurons in 3\textsuperscript{rd} HL & 64 & 64 & -- & 128 \\
        Activation Function & ReLU & ReLU & ReLU & ReLU \\
        Dropout Rate & 0.10 & 0.10 & 0.20 & 0.30 \\
        \midrule
        \textbf{Training Parameters} & & & & \\
        Optimizer & ADAM & ADAM & ADAM & ADAM \\
        Learning Rate & $5\times 10^{-4}$ & $5\times 10^{-4}$ & $1\times 10^{-3}$ & $1\times 10^{-4}$ \\
        Batch Size & 128 & 256 & 64 & 128 \\
        Number of Epochs & 100 & 100 & 30 & 50 \\
        Weight Decay & $1\times 10^{-5}$ & $1\times 10^{-5}$ & -- & $1\times 10^{-5}$ \\
        \midrule
        \textbf{Specific Parameters} & & & & \\
        Latent dimension & 128 & -- & -- & -- \\
        Constraint weight ($\phi$) & 0.5 & -- & -- & -- \\
        Temperature $\tau_{\text{cont}}$ & -- & 0.50 & -- & -- \\
        Temperature $\tau_{\text{cat}}$ & -- & 0.20 & -- & -- \\
        Replacement rate ($r$) & -- & 0.15 & -- & -- \\
        Head mixing ($\lambda$) & -- & 0.50 & -- & -- \\
        Projection dim (per head) & -- & 128 & -- & -- \\
        Refresh interval (epochs) & -- & 10 & -- & -- \\
        $F$-folds (CL OOS probs) & -- & -- & 5 & -- \\
        Quantile ($q$) & -- & -- & 0.70 & -- \\
        Minimum weight ($w_{\min}$) & -- & -- & 0.20 & -- \\
        Logistic slope ($\gamma$) & -- & -- & 4 & -- \\
        $\alpha$ (CE term) & -- & -- & -- & 0.3 \\
        $\beta$ (RCE term) & -- & -- & -- & 2.0 \\
        \bottomrule
    \end{tabular}
    }
\end{table}

\begin{figure*}[!ht]
\centering
\begin{subfigure}{0.49\textwidth}
\centering
\includegraphics[width=\columnwidth]{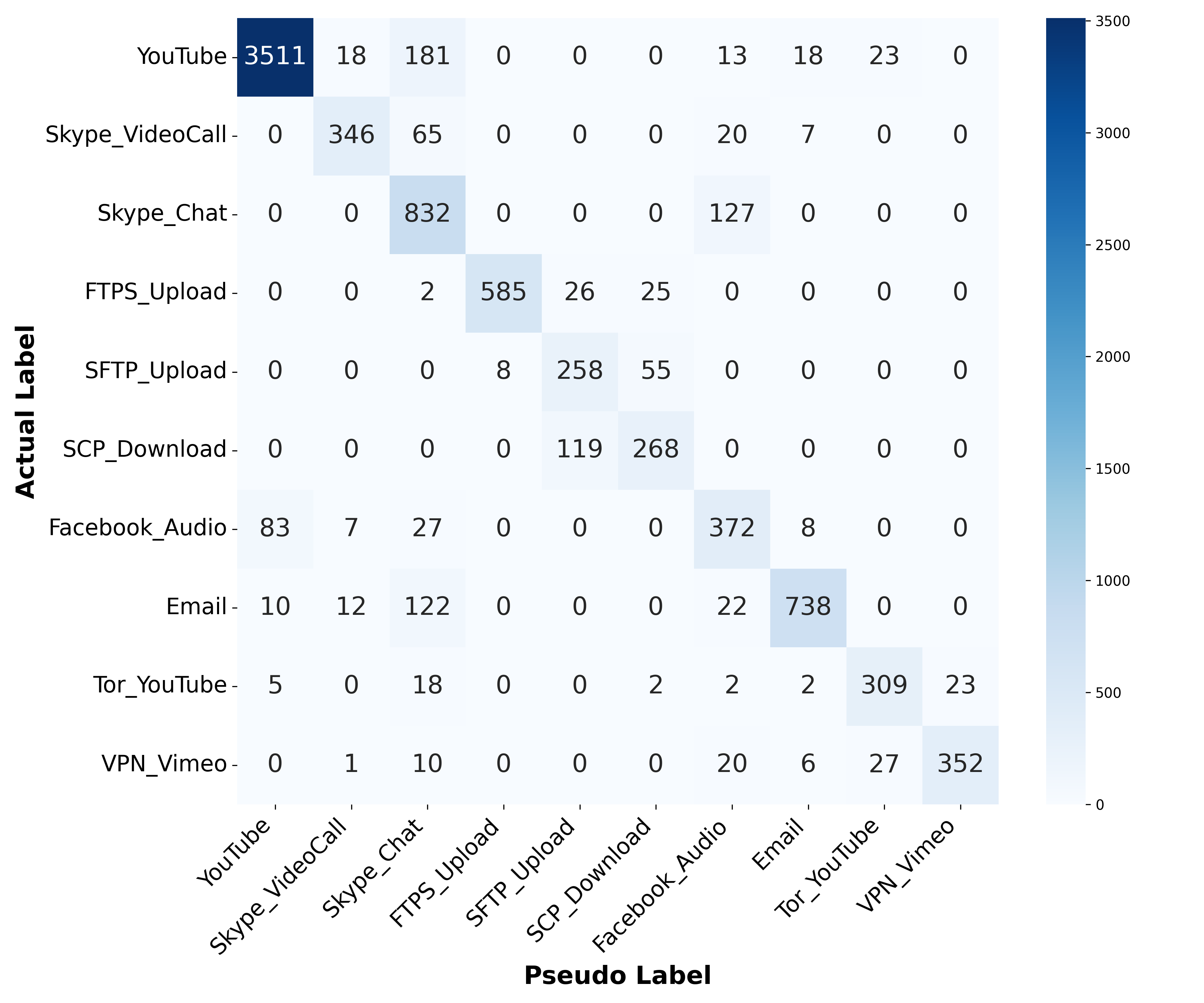}
\caption{\fontfamily{ptm}\fontsize{8}{10}\selectfont Autoencoder.}
\label{fig:autoencoder_confusion_matrix}
\end{subfigure}
\begin{subfigure}{0.49\textwidth}
\centering
\includegraphics[width=\columnwidth]{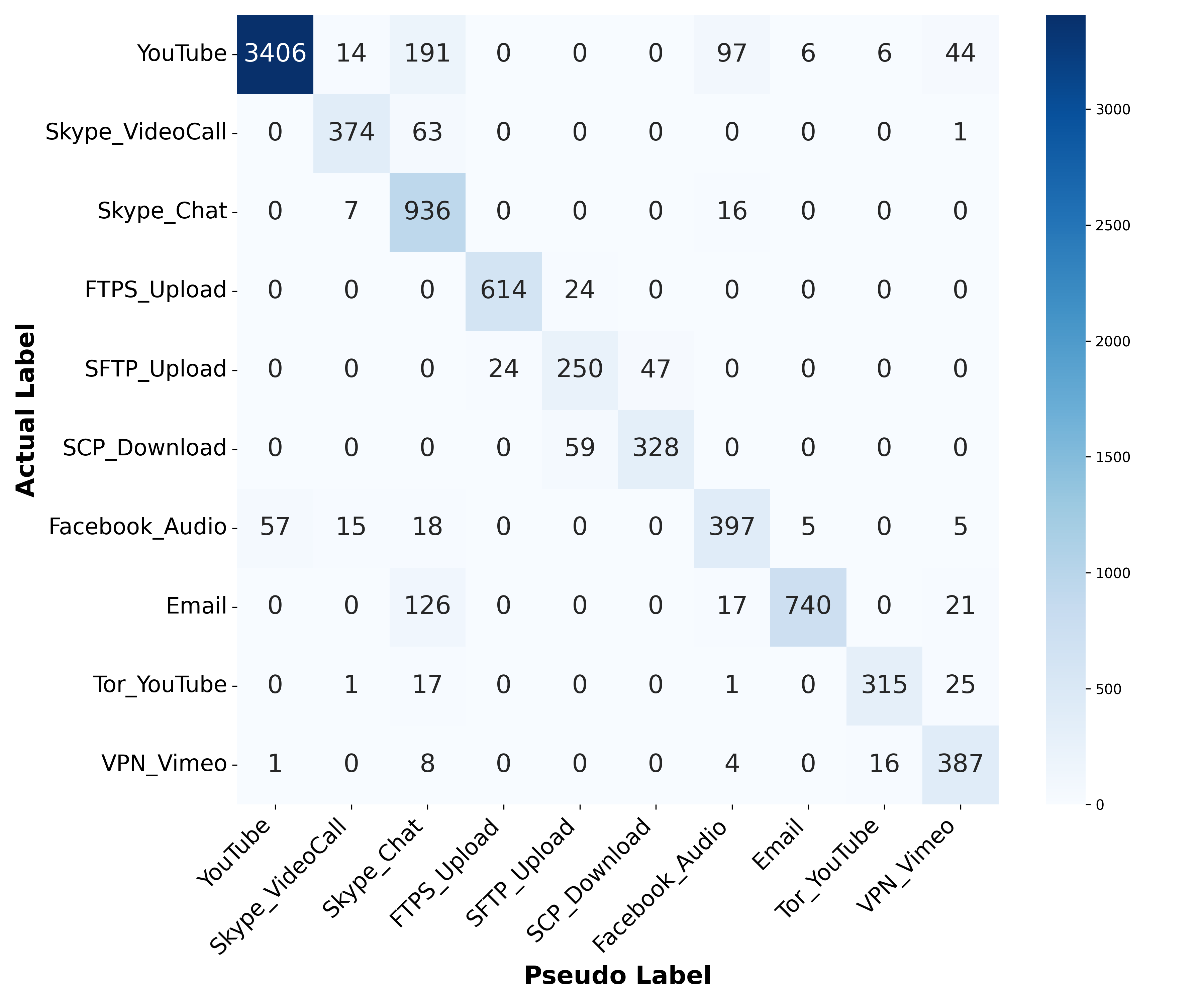}
\caption{\fontfamily{ptm}\fontsize{8}{10}\selectfont TabCL.}
\label{fig:tabcl_confusion_matrix}
\end{subfigure}
\caption{Confusion matrices of SSL models for adding pseudo-labels: (a) Autoencoder, (b) TabCL.}
\label{fig:confusion_matrices}
\end{figure*}

\subsubsection{Self-Supervised Learning for Pseudo-Label Generation}
Two SSL techniques were employed to assign pseudo-labels to the unlabeled dataset: Autoencoder and TabCL. These methods were applied separately as pretraining strategies on the unlabeled corpus (8684 flows), followed by fine-tuning with the small labeled set. The pseudo-label accuracy is 87.17\% for AE and 89.22\% for TabCL. We also evaluated a confidence–margin voting fusion that consolidates AE and TabCL predictions by agreement, this yields 89.91\% accuracy—an incremental gain over TabCL alone. In fusion, the improvement is modest because TabCL already provides strong representations and the two models’ errors partially overlap; however, fusion slightly increases precision, which is beneficial for the downstream confident-learning filter. Confusion matrices are presented in Fig. \ref{fig:confusion_matrices}, illustrating the alignment between pseudo-labels and true labels. Detailed metrics appear in Table~\ref{tab:performance_metrics}.

\begin{table}[h]
    \centering
    \caption{Performance metrics for SSL pseudo-labeling.}
    \label{tab:performance_metrics}
    \resizebox{\columnwidth}{!}{
    \begin{tabular}{lcccc}
        \toprule
        \textbf{Model} & \textbf{Precision} & \textbf{Recall} & \textbf{F1-score} & \textbf{Accuracy} \\
        \midrule
        Autoencoder & $88.30 \pm 0.3$ & $86.50 \pm 0.7$ & $87.20 \pm 0.3$ & $87.17 \pm 0.3$ \\
        TabCL & $90.10 \pm 0.5$ & $88.60 \pm 0.4$ & $88.90 \pm 0.2$ & $89.22 \pm 0.2$ \\
        (AE $\oplus$ TabCL) & $\mathbf{90.40 \pm 0.4}$ & $\mathbf{89.10 \pm 0.4}$ & $\mathbf{89.30 \pm 0.3}$ & $\mathbf{89.91 \pm 0.2}$ \\
        \bottomrule
    \end{tabular}
    }
\end{table}

Here, we report the computational footprint of the two SSL pre-training strategies. All values are the mean over 10 runs with different random seeds. As summarized in Table \ref{tab:ssl_compute}, the autoencoder (AE) is lightweight with 31.19 s per full training round and 1.2 GB peak VRAM, while TabCL trades additional compute for higher-quality representations: 107.86 s per round ($≈$3.5× AE) and 1.5 GB peak VRAM. The gap mainly comes from contrastive objectives (multi-view generation and similarity computations), which increase arithmetic intensity even with modest parameter counts. Importantly, both methods fit comfortably on a 15 GB T4 and keep peak memory below 2 GB, indicating they are deployable on commodity GPUs. The pseudo-labeling pass itself is not the bottleneck (AE: 93,464 flows/s; TabCL: 88,409 flows/s), so end-to-end cost is dominated by pre-training. In terms of scalability, measured throughput scales near-linearly with the number of flows, so larger corpora increase wall-clock time proportionally. Overall, the compute profile shows that our SSL stage is efficient in absolute terms (minutes per round) and scalable to larger traffic datasets while remaining resource-frugal.

\begin{table}[h]
    \centering
    \caption{Compute profile of SSL pseudo-label generation.}
    \label{tab:ssl_compute}
    \begin{tabular}{lcc} 
        \toprule
        \textbf{Metric} & \textbf{Autoencoder} & \textbf{TabCL} \\
        \midrule
Time per round (s)                        & 31.19 & 107.86 \\
Training throughput (flows/s)             & 27861  & 8092  \\
Pseudo-labeling throughput (flows/s)      & 93464  & 88409  \\
Peak GPU memory (GB)                      & 1.2   & 1.5   \\
Average GPU utilization (\%)              & 32.2    & 43.1    \\
Trainable parameters (M)                     & 0.032   & 0.069   \\
        \bottomrule
    \end{tabular}
\end{table}

\subsubsection{Final Classifier with Confident Learning}
\begin{figure}[h]
    \centering
    \includegraphics[width=\columnwidth]{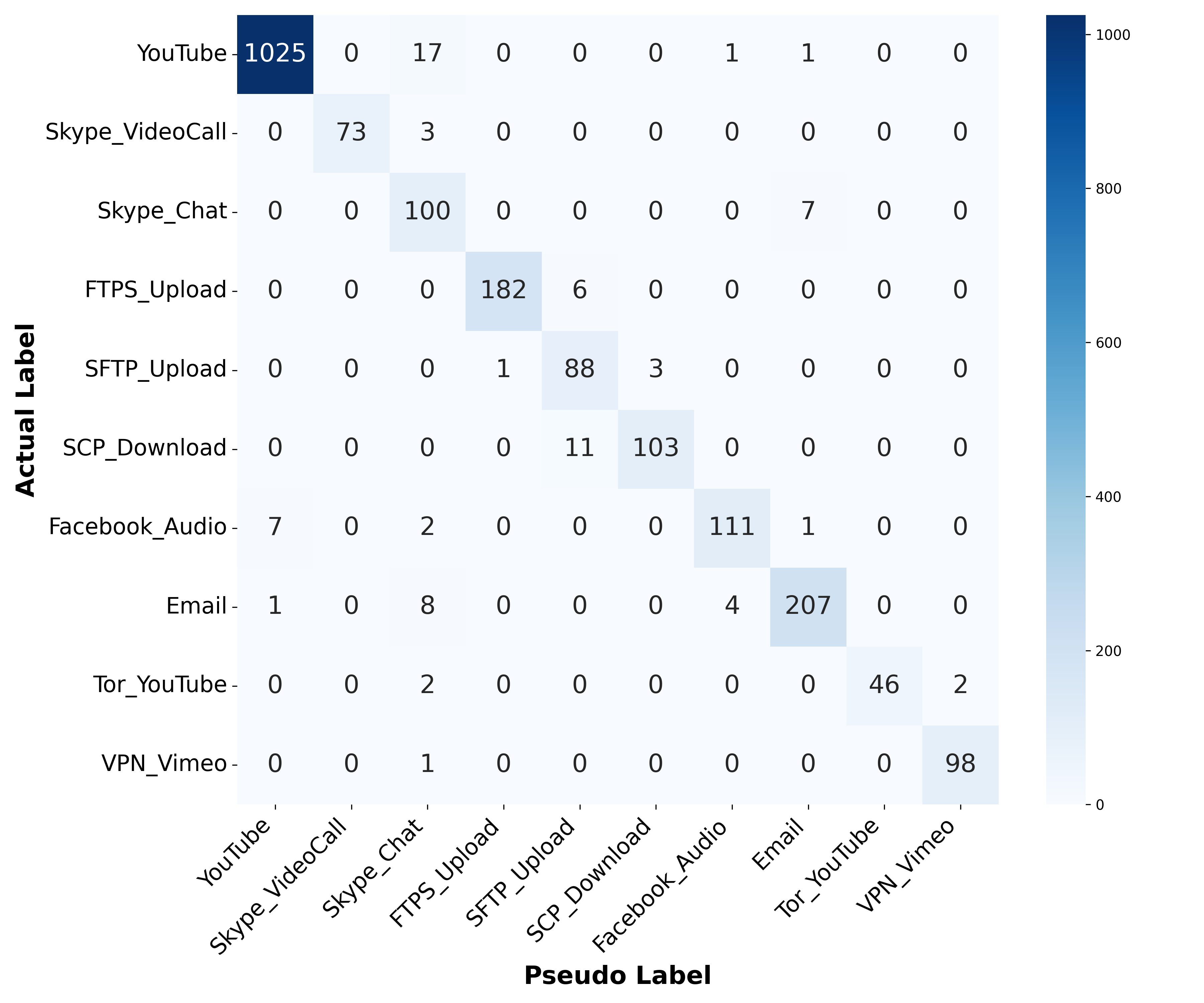}
    \caption{Confusion matrix for the MLP final classifier on the test dataset.}
    \label{fig:classifier_confusion_matrix}
\end{figure}

\begin{figure}[h]
    \centering
    \includegraphics[width=\columnwidth]{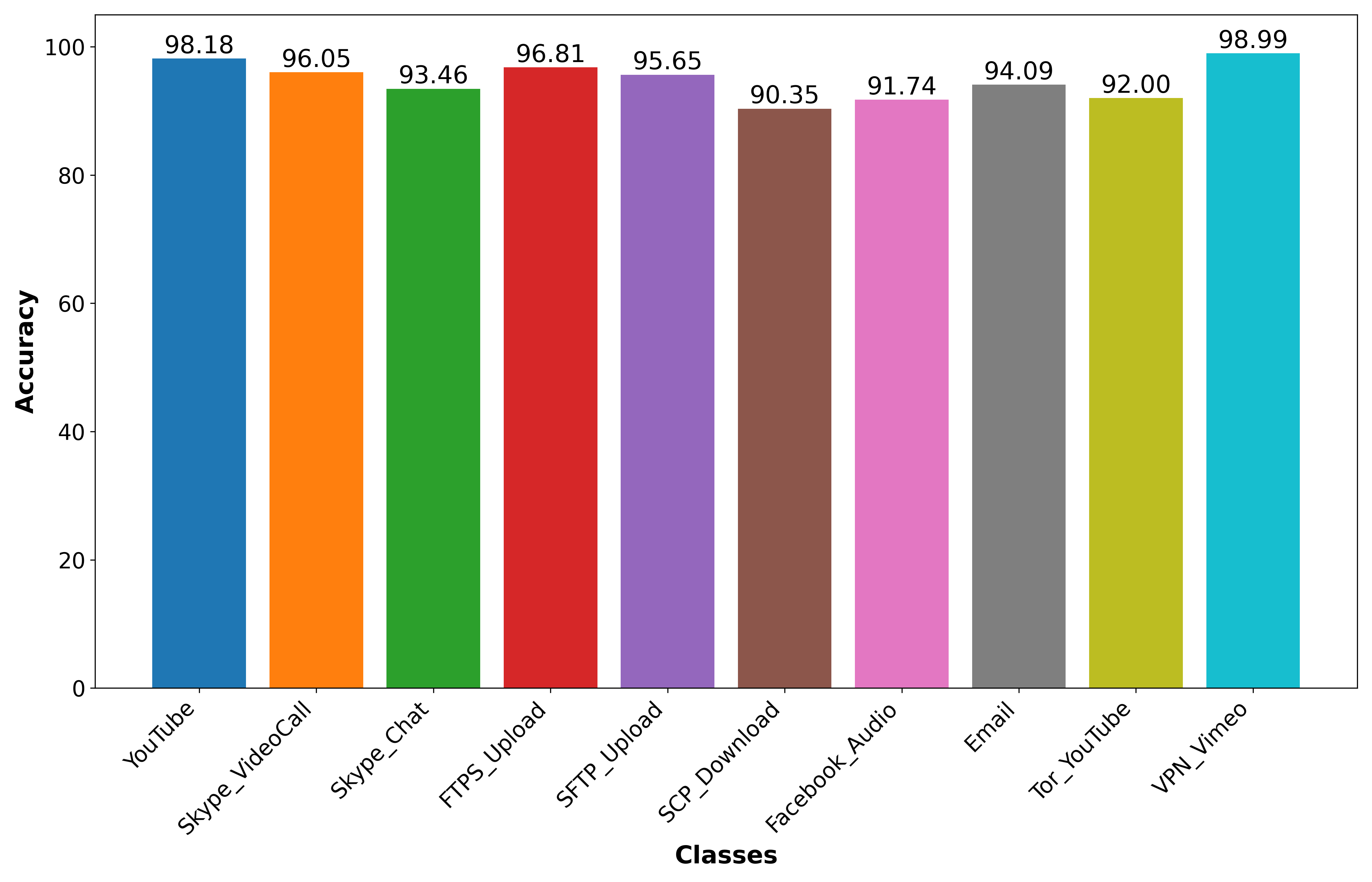}
    \caption{Per-class accuracy of final classifier.}
    \label{fig:class_wise_accuracy}
\end{figure}

For training the final classifier, we use the fused pseudo-labels produced by the confidence–margin voting of AE and TabCL from the previous subsection (Fig.~\ref{fig:confusion_matrices}, Table~\ref{tab:performance_metrics}). The fused pool attains an overall pseudo-label accuracy of $89.91\%$, implying that roughly $10\%$ of labels are noisy. If this noise is not addressed and the final multilayer perceptron (MLP) is trained directly on the fused pseudo-labels, the test accuracy plateaus at $90.9\%$. To improve robustness, we apply traffic-adopted confident learning to estimate per-sample label reliability and downweight likely errors before final training. With CL, the final MLP reaches an accuracy of $\mathbf{96.29\%}$. The confusion matrix for the final classifier is shown in Fig.~\ref{fig:classifier_confusion_matrix}, and per-class metrics are reported in Table~\ref{tab:class_performance_from_cm}, which shows uniformly high scores across most classes, with the few lower metrics because of feature similarity between these types of traffics. Additionally, a bar chart in Fig. \ref{fig:class_wise_accuracy} illustrates the accuracy of the final classifier across individual classes, indicating that class-imbalance effects were effectively mitigated by the traffic-adopted CL stage.

\begin{table}[h]
    \centering
    \caption{Per-class performance of the final classifier.}
    \label{tab:class_performance_from_cm}
    \resizebox{\columnwidth}{!}{
    \begin{tabular}{lccc}
        \toprule
        \textbf{Class} & \textbf{Precision (\%)} & \textbf{Recall (\%)} & \textbf{F1 (\%)} \\
        \midrule
        YouTube          & 99.23 & 98.18 & 98.70 \\
        Skype\_VideoCall & 100.00 & 96.05 & 97.99 \\
        Skype\_Chat      & 75.19 & 93.46 & 83.33 \\
        FTPS\_Upload     & 99.45 & 96.81 & 98.11 \\
        SFTP\_Upload     & 83.81 & 95.65 & 89.34 \\
        SCP\_Download    & 97.17 & 90.35 & 93.64 \\
        Facebook\_Audio  & 95.69 & 91.74 & 93.67 \\
        Email            & 95.83 & 94.09 & 94.95 \\
        Tor\_YouTube     & 100.00 & 92.00 & 95.83 \\
        VPN\_Vimeo       & 98.00 & 98.99 & 98.49 \\
        \midrule
        \textbf{Overall (weighted avg)} & \textbf{96.68} & \textbf{96.31} & \textbf{96.41} \\
        \bottomrule
    \end{tabular}
    }
\end{table}

Table~\ref{tab:ablation_cl} summarizes a comprehensive ablation across three axes. (i) Pseudo-label source: keeping the downstream CL+MLP stage fixed, we vary the source of pseudo-labels (AE, TabCL, or their fusion). Fusion delivers consistent gains over either branch alone, indicating that higher-quality, lower-noise pseudo-labels translate directly into stronger downstream performance. 
(ii) CL variants: fixing the pseudo-label source to Fusion, we progressively enrich CL from a basic mean-threshold filter to our traffic-adapted pipeline (quantile thresholds, logistic weighting, and BRC). Each addition yields measurable improvements, demonstrating that the traffic-aware CL design outperforms the original CL heuristic. 
(iii) Component contribution: isolating components shows that either SSL (without CL) or CL (without SSL pretraining, just pseudo labeling with small labeled dataset) is insufficient; the full pipeline (SSL + CL) is required to reach the best accuracy and macro-F1 with a small labeled dataset.

\begin{table}[h]
    \centering
    \caption{Ablation of the final stage.}
    \label{tab:ablation_cl}
    \resizebox{\columnwidth}{!}{
    \begin{tabular}{lcc}
        \toprule
        \textbf{Setting} & \textbf{Macro-F1 (\%)} & \textbf{Accuracy (\%)} \\
        \midrule
        \multicolumn{3}{l}{\textbf{Pseudo-label source (for CL+MLP stage)}} \\
        \quad AE pseudo-labels + CL              & $94.3 \pm 0.4$ & $94.2 \pm 0.4$ \\
        \quad TabCL pseudo-labels + CL           & $95.7 \pm 0.3$ & $95.6 \pm 0.3$ \\
        \quad \textbf{Fusion (AE $\oplus$ TabCL) + CL} & $\mathbf{96.43 \pm 0.3}$ & $\mathbf{96.29 \pm 0.3}$ \\
        \midrule
        \multicolumn{3}{l}{\textbf{CL variants (on Fusion pseudo-labels)}} \\
        \quad No-CL (direct training)            & $91.0 \pm 0.5$ & $90.9 \pm 0.5$ \\
        \quad Mean threshold (original CL)       & $95.0 \pm 0.4$ & $95.2 \pm 0.4$ \\
        \quad Quantile threshold only            & $95.4 \pm 0.3$ & $95.4 \pm 0.3$ \\
        \quad + Logistic weighting               & $96.0 \pm 0.2$ & $95.9 \pm 0.2$ \\
        \quad + BRC (full CL)                    & $\mathbf{96.43 \pm 0.3}$ & $\mathbf{96.29 \pm 0.3}$ \\
        \midrule
        \multicolumn{3}{l}{\textbf{Component contribution}} \\
        \quad Only SSL (Fusion, no CL)           & $91.0 \pm 0.5$ & $90.9 \pm 0.5$ \\
        \quad Only CL (no SSL pretraining)       & $85.7 \pm 0.6$ & $85.8 \pm 0.7$ \\
        \quad SSL + CL (ours)                    & $\mathbf{96.43 \pm 0.3}$ & $\mathbf{96.29 \pm 0.3}$ \\
        \bottomrule
    \end{tabular}
    }
\end{table}

\begin{table*}[h]
    \centering
    \caption{Performance Metrics of our model and state-of-the-art methods}
    \label{tab:state-of-the-art}

    \begin{tabular}{lccccc} 
        \toprule
        \textbf{Methods} & \textbf{Authors \& Years} & \textbf{Precision} & \textbf{Recall} & \textbf{F1-score} & \textbf{Accuracy} \\ 
        \midrule
        Multitask learning fusion & L. Liu et al. 2024 \cite{35} & 90.24 & 91.2 & 90.05 & 92.76 \\
        TFE-GNN & H. Zhang et al. 2023 \cite{45} & 94.21 & 93.91 & 93.88 & 93.45 \\
        SSL FlowPic & E. Horowicz et al. 2024 \cite{18} & 92.80 & 95.22 & 93.5 & 93.73 \\
        ET-BERT (flow-level) & X. Lin et al. 2022 \cite{19} & 93.40 & 94.01 & 93.49 & 91.64 \\
        Encrypted Traffic classification with SSL & M. Towhid et al. 2023 \cite{21} & 88.4 & 89.02 & 87.56 & 90.56 \\
        \midrule
        \textbf{Our Method} & Ours & \textbf{96.68} & \textbf{96.31} & \textbf{96.41} & \textbf{96.29} \\  
        \bottomrule
    \end{tabular}
\end{table*}

\begin{figure}[h]
    \centering
    \includegraphics[width=\columnwidth]{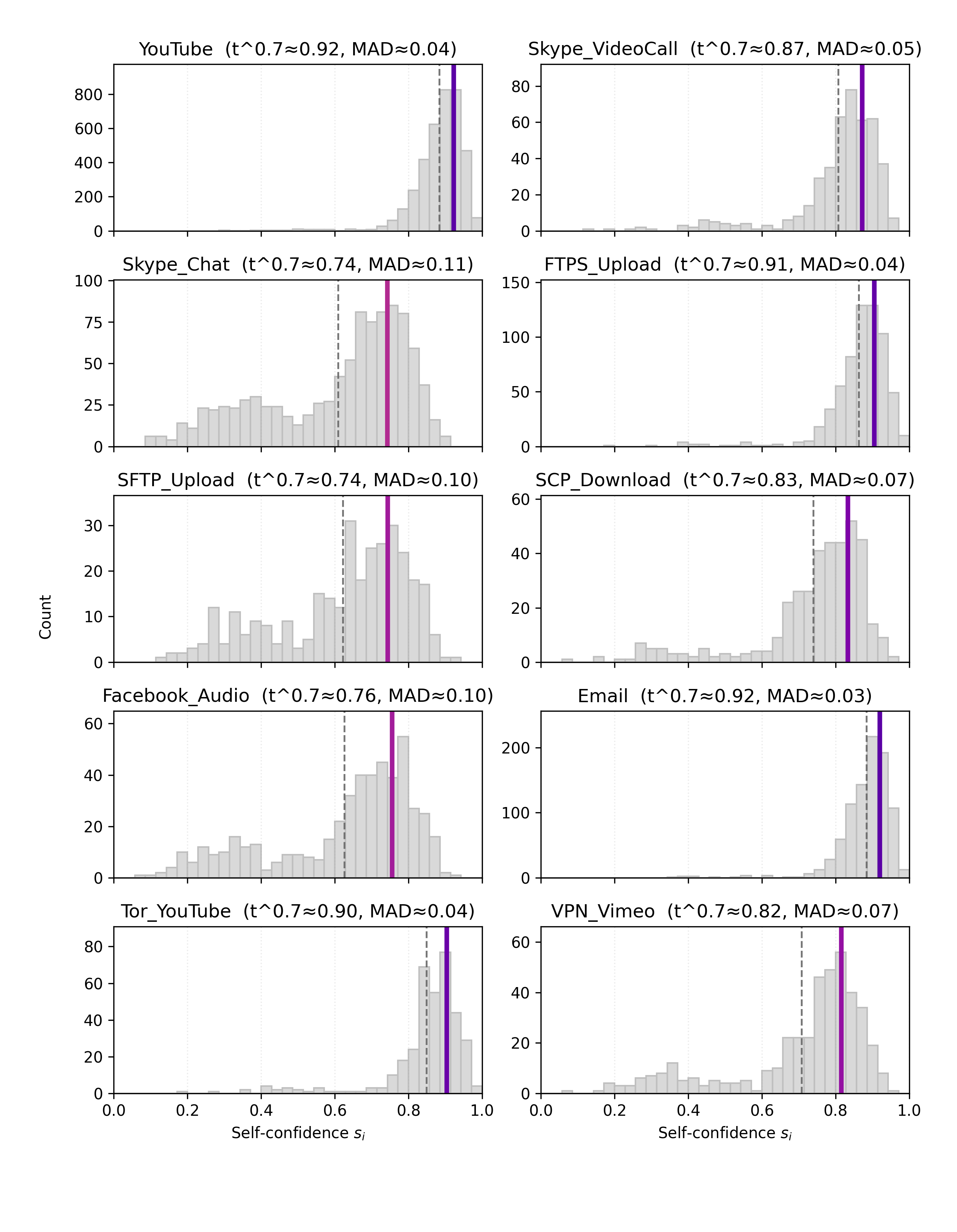}
    \caption{Per-class distributions of the self-confidence $s_i$ for all pseudo-labeled flows.}
    \label{fig:cl_thresholds}
\end{figure}

Fig.~\ref{fig:cl_thresholds} visualizes how CL adapts its per-class decision thresholds to the reliability of the pseudo-labels. The dashed gray line is the class mean (original CL), and the magenta solid line is the quantile threshold $t^{0.7}_j$. Clean classes in the SSL output (e.g., YouTube, FTPS\_Upload, and Email) exhibit sharply peaked histograms near $s_i\!\approx\!0.9$--$1.0$, small MAD ($\approx\!0.03$--$0.05$), and correspondingly high thresholds $t^{0.7}_j\!\in\![0.90,0.92]$. In contrast, classes that were noisier in the SSL (e.g., Skype\_Chat, SFTP\_Upload, and Facebook\_Audio) show broader, left-shifted distributions with larger MAD ($\approx\!0.07$--$0.11$) and noticeably lower thresholds, with $t^{0.7}_j$ moving to $0.74$--$0.82$. This leftward adjustment is intentional: it prevents aggressive pruning of inherently harder classes by requiring a less stringent confidence to contribute significant weight. The dashed gray mean lines are often left of the magenta $t^{0.7}_j$ in clean classes due to right-skew; using the quantile rather than the mean avoids being unduly influenced by a small tail of low-confidence samples. Conversely, in noisy classes the quantile naturally tracks lower, reflecting the heavier mass at moderate confidences. These adaptive thresholds feed our CL weighting. As a result, the final classifier trained with CL improves from $90.9\%$ (no CL) to $96.29\%$ accuracy. The largest per-class gains occur exactly where Fig.~\ref{fig:cl_thresholds} indicates higher noise and lower $t^{0.7}_j$ (e.g., Skype\_Chat, SFTP\_Upload, and Facebook\_Audio), confirming that the performance gap between SSL pseudo-labeling and the final model is attributable to the CL stage.

\begin{figure*}[!ht]
\centering
\begin{subfigure}{0.49\textwidth}
\centering
\includegraphics[width=\columnwidth]{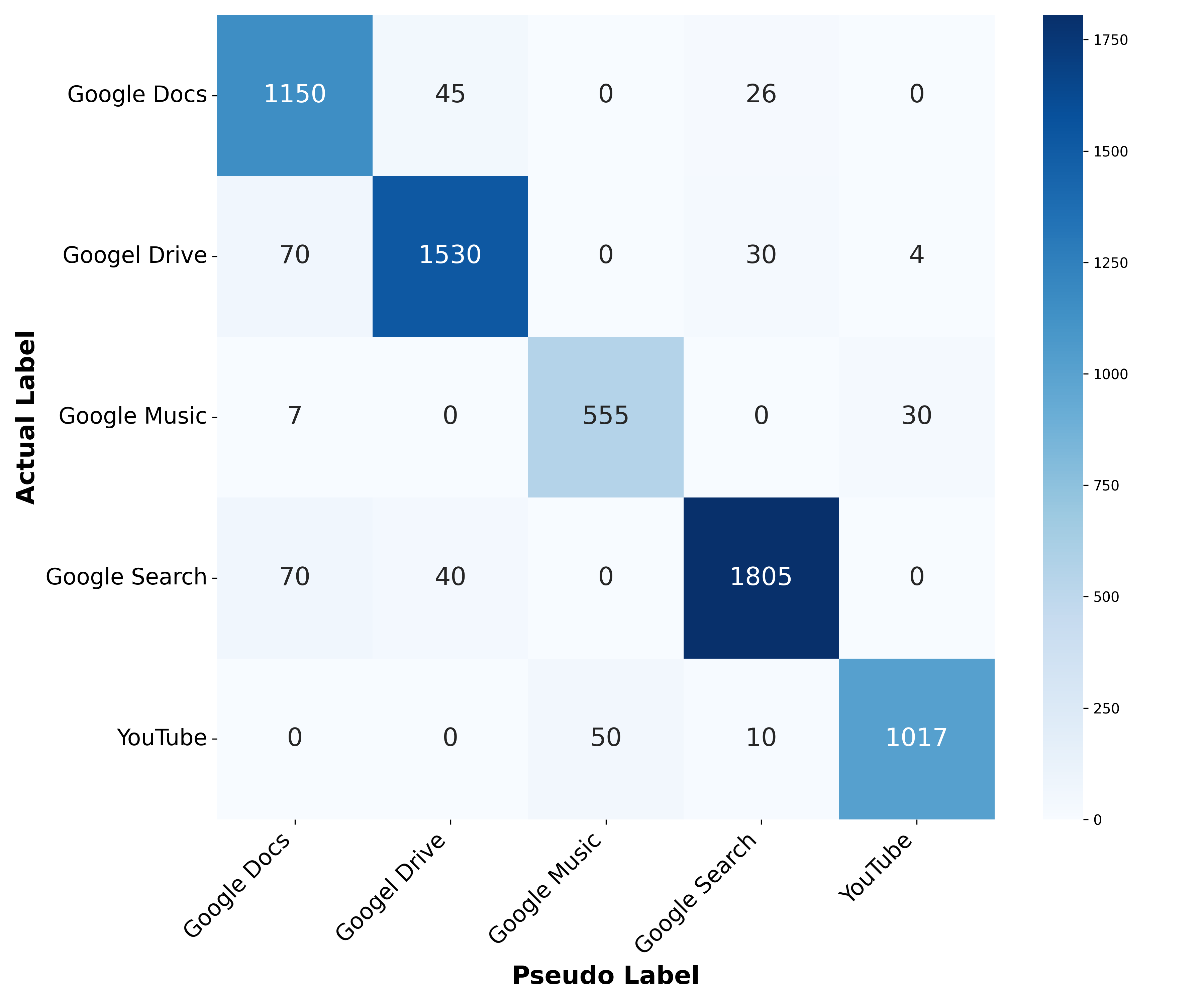}
\caption{\fontfamily{ptm}\fontsize{8}{10}\selectfont Autoencoder (AE).}
\end{subfigure}
\begin{subfigure}{0.49\textwidth}
\centering
\includegraphics[width=\columnwidth]{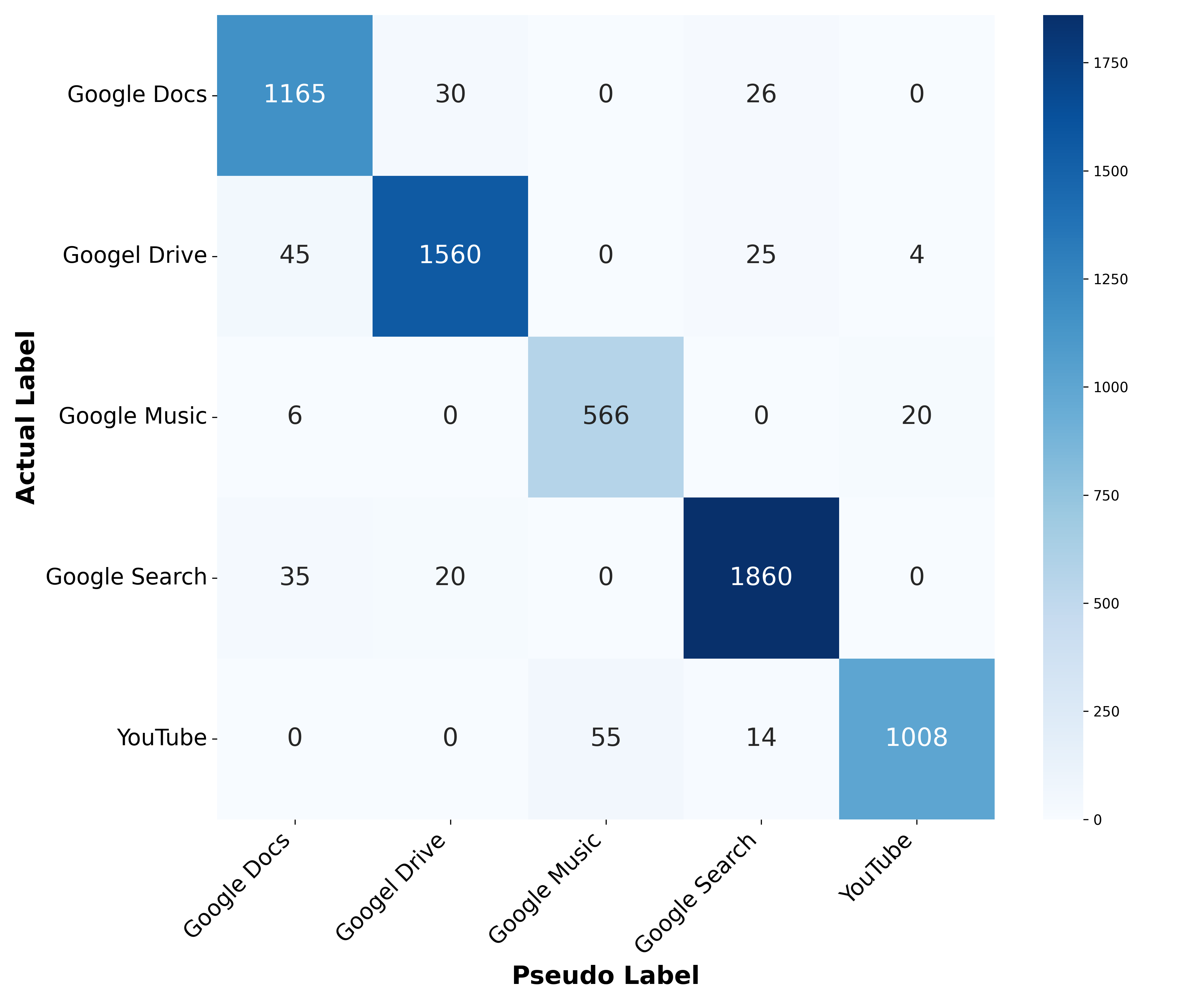}
\caption{\fontfamily{ptm}\fontsize{8}{10}\selectfont TabCL.}
\end{subfigure}
\caption{UCDavis--QUIC: confusion matrices of SSL pseudo-labelers.}
\label{fig:quic_ssl_cm}
\end{figure*}

To further validate our approach, we compare our framework with five recent approaches that span supervised deep models and SSL paradigms: an MTEFU model that combines CNN/SAE/GRU/LSTM heads under a shared architecture (supervised) \cite{35}, TFE-GNN (supervised, graph neural model) \cite{45}, FlowPic with self-supervised contrastive pre-training on mini-FlowPic images \cite{18}, ET-BERT \cite{19} with transformer unsupervised pre‑training followed by supervised fine‑tuning on limited labels at packet/flow level, and an SSL approach for encrypted traffic \cite{21}. To make the comparison fair and reproducible, we re‑implemented (or followed the authors’ public descriptions) and evaluated all baselines under our datasets. Specifically, methods that require labels only (TFE‑GNN and multitask fusion) were trained on the self-generated dataset and evaluated on the same test partitions as ours; methods that require unlabeled + a small labeled set (FlowPic, ET‑BERT, Towhid \& Shahriar) were provided our self‑generated corpus as unlabeled plus the same small labeled split, mirroring their intended usage.

As summarized in Table \ref{tab:state-of-the-art}, our framework achieves 96.29\% accuracy while using only a small labeled seed, outperforming FlowPic (93.73\%) and ET-BERT (91.64\%)under our setting, and exceeding the supervised TFE‑GNN and multitask‑fusion baselines despite their access to labeled training data. Despite those powerful pre-training strategies, our SSL + CL pipeline delivers higher accuracy with far fewer labels, indicating that flow-tabular representation learning combined with noise-aware confident learning is competitive with image- and transformer-based pre-training while being label-efficient. Beyond the headline accuracy, our framework operates end-to-end—from raw packet capture to flow aggregation, pseudo-labeling, noise filtering, and final classification—so it can be deployed in realistic “low-label” settings rather than only curated benchmarks. Evaluations on our controlled, end-to-end corpus and on the public ISCX subset indicate that the method scales to thousands of flows and transfers with minimal tuning, suggesting good generalization while directly addressing label scarcity through SSL and CL.

\subsection{Performance of the Proposed Model on UCDavis--QUIC}
\label{subsec:quic_results}
To stress--test generalization beyond our self-generated/ISCX setting, we further evaluate on the UCDavis--QUIC dataset with five Google services (Docs, Drive, Music, Search, YouTube). In keeping with our ``low-label'' protocol, we treat \textbf{5\%} of flows per class as the small labeled set and the remaining \textbf{95\%} as unlabeled (Table~\ref{tab:quic_class_dist}). Hyperparameters are selected exactly as in the previous subsection (stratified 5-fold CV on the small labeled split; three seeds per candidate; select by Macro-F1). Given the lower number of classes (5) and a more homogeneous feature schema, only minor changes were adopted: based on grid search, for the AE we use a slightly smaller latent size (64) to curb overfitting; in TabCL we reduce augmentation intensity (replacement rate $r\!=\!0.10$) and temperature ($\tau_{\text{cont}}\!=\!0.4$, $\tau_{\text{cat}}\!=\!0.15$) to reflect the smoother intra-class structure; in CL we raise the quantile threshold to $q\!=\!0.75$ and use a steeper logistic slope $\gamma\!=\!6$ to be more selective with residual noise.

\subsubsection{Self-Supervised Learning for Pseudo-Label Generation (QUIC)}
We apply both SSL branches independently on the unlabeled QUIC flows and then fine-tune on the 5\% labeled split. Pseudo-label performance on the full dataset is strong: the AE reaches $94.07\%$ accuracy, TabCL improves to $95.65\%$, and a confidence--margin voting fusion gives a further lift to $96.11\%$. Class-wise confusion patterns for AE and TabCL are shown in Fig.~\ref{fig:quic_ssl_cm}. Aggregate metrics of SSL pseudo labeling are detailed in Table~\ref{tab:quic_ssl_metrics}.

\begin{table}[h]
    \centering
    \caption{UCDavis--QUIC: SSL pseudo-labeling metrics.}
    \label{tab:quic_ssl_metrics}
    \resizebox{\columnwidth}{!}{
    \begin{tabular}{lcccc}
        \toprule
        \textbf{Model} & \textbf{Precision (\%)} & \textbf{Recall (\%)} & \textbf{F1 (\%)} & \textbf{Accuracy (\%)} \\
        \midrule
        Autoencoder & $94.16 \pm 0.28$ & $94.07 \pm 0.26$ & $94.09 \pm 0.24$ & $94.07 \pm 0.22$ \\
        TabCL & $95.70 \pm 0.22$ & $95.65 \pm 0.21$ & $95.66 \pm 0.21$ & $95.65 \pm 0.20$ \\
        (AE $\oplus$ TabCL) & $\mathbf{96.20 \pm 0.20}$ & $\mathbf{96.10 \pm 0.20}$ & $\mathbf{96.15 \pm 0.19}$ & $\mathbf{96.11 \pm 0.18}$ \\
        \bottomrule
    \end{tabular}
    }
\end{table}

For completeness, we profiled the SSL round times on the same workstation used earlier. Owing to the smaller corpus (6{,}439 flows vs.\ 8{,}684 previously), both branches are faster: AE averages $22.21$\,s/round and TabCL $51.95$\,s/round. Peak GPU memory is modest for both branches—about 1.0\,GB for AE and 1.3\,GB for TabCL. This is consistent with the roughly linear scaling of throughput with the number of flows observed in the previous subsection.

\subsubsection{Final Classifier with Confident Learning (QUIC)}
\begin{figure}[h]
    \centering
    \includegraphics[width=\columnwidth]{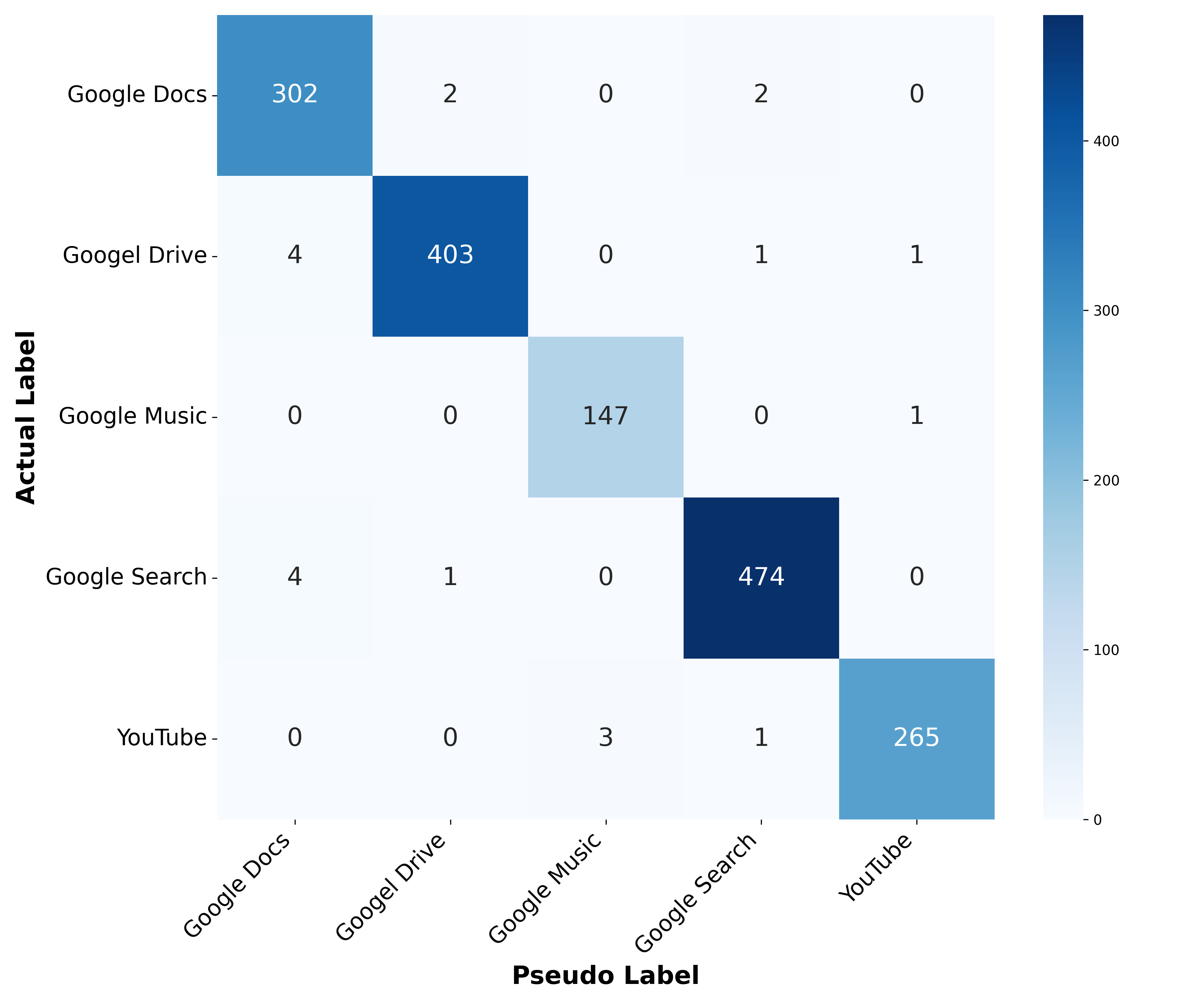}
    \caption{UCDavis--QUIC: confusion matrix of the final MLP with CL.}
    \label{fig:quic_final_cm}
\end{figure}

We then pass the fused pseudo-labeled pool into traffic-adapted CL and train the final MLP. The resulting model attains \textbf{$98.76\% \pm 0.10$} accuracy. The confusion matrix is shown in Fig.~\ref{fig:quic_final_cm}. Per-class precision, recall, F1, and per-class accuracy are reported in Table~\ref{tab:quic_perclass}. Scores are uniformly high; the few residual errors arise primarily from semantically similar services (e.g., Music vs.\ YouTube, Docs vs.\ Search).

\begin{table}[h]
    \centering
    \caption{UCDavis--QUIC: per-class performance of the final classifier.}
    \label{tab:quic_perclass}
    \resizebox{\columnwidth}{!}{
    \begin{tabular}{lcccc}
        \toprule
        \textbf{Class} & \textbf{Precision (\%)} & \textbf{Recall (\%)} & \textbf{F1 (\%)} & \textbf{Accuracy (\%)} \\
        \midrule
        Google Docs   & 97.49 & 98.77 & 98.09 & 98.69 \\
        Google Drive  & 99.20 & 98.53 & 98.90 & 98.59 \\
        Google Music  & 97.67 & 98.99 & 98.32 & 98.99 \\
        Google Search & 99.22 & 98.96 & 99.14 & 99.06 \\
        YouTube       & 99.25 & 98.51 & 98.79 & 98.33 \\
        \midrule
        \textbf{Overall (weighted)} & \textbf{98.85} & \textbf{98.74} & \textbf{98.69} & \textbf{98.76} \\
        \bottomrule
    \end{tabular}
    }
\end{table}

Table~\ref{tab:ablation_quic} consolidates three ablations using the same protocol as before (mean over 10 seeds). (i) Pseudo-label source. Holding the downstream CL+MLP fixed, (ii) CL variants Starting from ``No-CL'', progressively adding quantile thresholds, logistic weighting, and our BRC step yields monotonic gains—evidence that the traffic-adapted CL is more effective than the original mean-threshold heuristic. (iii) Component contribution Neither SSL (without CL) nor CL (without SSL pretraining) is sufficient; the full SSL+CL pipeline is required to reach the best accuracy/F1 with only 5\% labels.

\begin{table}[h]
    \centering
    \caption{Ablation on UCDavis--QUIC.}
    \label{tab:ablation_quic}
    \resizebox{\columnwidth}{!}{
    \begin{tabular}{lcc}
        \toprule
        \textbf{Setting} & \textbf{Macro-F1 (\%)} & \textbf{Accuracy (\%)} \\
        \midrule
        \multicolumn{3}{l}{\textbf{Pseudo-label source (for CL+MLP)}} \\
        \quad AE pseudo-labels + CL              & $97.33 \pm 0.15$ & $97.30 \pm 0.15$ \\
        \quad TabCL pseudo-labels + CL           & $98.11 \pm 0.12$ & $98.05 \pm 0.12$ \\
        \quad \textbf{Fusion (AE $\oplus$ TabCL) + CL} & $\mathbf{98.85 \pm 0.10}$ & $\mathbf{98.76 \pm 0.10}$ \\
        \midrule
        \multicolumn{3}{l}{\textbf{CL variants (Fusion pseudo-labels)}} \\
        \quad No-CL (direct training)            & $96.22 \pm 0.18$ & $96.19 \pm 0.18$ \\
        \quad Mean threshold (original CL)       & $98.31 \pm 0.12$ & $98.25 \pm 0.12$ \\
        \quad Quantile threshold only            & $98.47 \pm 0.12$ & $98.42 \pm 0.12$ \\
        \quad + Logistic weighting               & $98.66 \pm 0.11$ & $98.62 \pm 0.11$ \\
        \quad + BRC (full CL)                    & $\mathbf{98.85 \pm 0.10}$ & $\mathbf{98.76 \pm 0.10}$ \\
        \midrule
        \multicolumn{3}{l}{\textbf{Component contribution}} \\
        \quad Only SSL (Fusion, no CL)           & $96.24 \pm 0.18$ & $96.18 \pm 0.18$ \\
        \quad Only CL (no SSL pretraining)       & $93.00 \pm 0.22$ & $92.90 \pm 0.27$ \\
        \quad SSL + CL (ours)                    & $\mathbf{98.85 \pm 0.10}$ & $\mathbf{98.76 \pm 0.10}$ \\
        \bottomrule
    \end{tabular}
    }
\end{table}

\begin{table*}[h]
    \centering
    \caption{UCDavis--QUIC: comparison our model with state-of-the-art methods.}
    \label{tab:quic_sota}
    \begin{tabular}{lccccc}
        \toprule
        \textbf{Methods} & \textbf{Authors \& Years} & \textbf{Precision} & \textbf{Recall} & \textbf{F1-score} & \textbf{Accuracy} \\
        \midrule
        Multitask learning fusion & L. Liu et al., 2024 \cite{35} & 94.80 & 94.58 & 94.69 & 94.67 \\
        SSL FlowPic (QUIC) & E. Horowicz et al., 2024 \cite{18} & 98.76 & 99.66 & 98.71 & 98.30 \\
        Encrypted Traffic classification (SSL) & M. Towhid et al., 2023 \cite{21} & 98.12 & 97.93 & 98.03 & 98.01 \\
        \midrule
        \textbf{Our Method (SSL + CL)} & \textbf{Ours} & \textbf{98.85} & \textbf{98.74} & \textbf{98.69} & \textbf{98.76} \\
        \bottomrule
    \end{tabular}
\end{table*}

Finally, Table~\ref{tab:quic_sota} contrasts our method with three representative baselines that report results on UCDavis--QUIC. FlowPic’s QUIC “script” setting reports $98.3\%$ accuracy; the multitask fusion model reports $\approx\!94.7\%$ for ``Class'' prediction; the SSL method of Towhid \& Shahriar reports $\approx\!98\%$ depending on the pretraining ratio. As seen, our SSL\,+\,CL pipeline attains the strongest overall scores on this dataset as well, reinforcing that the framework generalizes across three datasets (in two evaluation settings) and remains competitive even against specialized supervised or SSL baselines.

\section{Conclusion}
\label{sec:conclusion} 
This paper presented an end-to-end framework for network traffic classification that couples self-supervised learning (SSL) with a traffic-adapted confident learning (CL) stage and a final weighted MLP classifier. Two complementary SSL branches, an autoencoder (AE) with constraint-consistent reconstruction and a tabular contrastive learner (TabCL) with class-conditioned views and dual-head temperatures, produce high-quality pseudo-labels from unlabeled flows. A confidence–margin voting fusion consolidates the branches, and the downstream CL module denoises pseudo-labels via per-class quantile thresholds, calibration-aware logistic weighting, and balanced retention before final training with symmetric cross-entropy. Across three datasets and two settings, the framework is accurate and label-efficient. On our deployment-style pipeline built from raw packet capture (self-generated unlabeled corpus) and a small ISCX VPN–nonVPN labeled split, the final model attains \textbf{96.29\%} accuracy with strong macro-F1, while remaining resource-frugal. On the public UCDavis–QUIC dataset, using only \textbf{5\%} labels per class, the method reaches \textbf{98.76\%} accuracy and uniformly high per-class scores, outperforming strong supervised and SSL baselines reported on the same data. Ablations confirm that (i) higher-quality pseudo-labels (AE/TabCL fusion) translate into better downstream performance, and (ii) the traffic-adapted CL design materially improves over a basic mean-threshold heuristic.

Beyond headline metrics, the framework operates end-to-end from raw packets to flow features, pseudo-labeling, noise mitigation, and classification—supporting realistic “low-label” deployments. Future work includes open-world/unknown-class handling, online domain adaptation under drift, and federated training for privacy-preserving scenarios.

\ifCLASSOPTIONcaptionsoff
  \newpage
\fi


%

\begin{IEEEbiography}[{\includegraphics[width=1in,height=1.25in,clip,keepaspectratio]{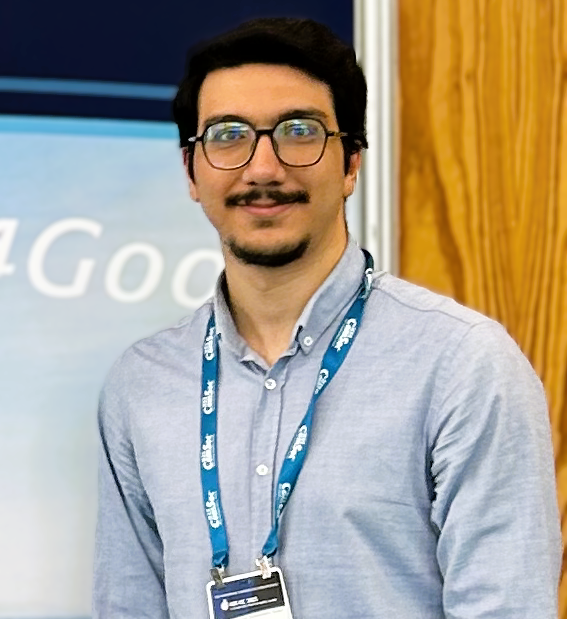}}]{Ehsan Eslami } (Member, IEEE) received the B.Sc. degree in computer engineering from the University of Tabriz, Tabriz, Iran, in 2022. He is currently pursuing his M.Sc. degree in computer engineering from Concordia University, Montreal, QC, Canada.

His research interests include AI-powered computer networks, federated learning, programmability of computer networks, and software-defined networking.
\end{IEEEbiography}

\begin{IEEEbiography}[{\includegraphics[width=1in,height=1.25in,clip,keepaspectratio]{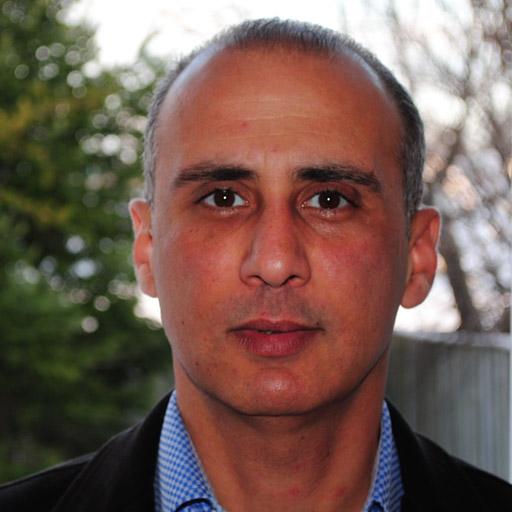}}]{Walaa Hamouda } ((Senior Member, IEEE) received the M.A.Sc. and Ph.D. degrees in electrical and computer engineering from Queens University, Kingston, ON, Canada, in 1998 and 2002, respectively. Since July 2002, he has been with the Department of Electrical and Computer Engineering, Concordia University, Montreal, QC, Canada, where he is currently a Professor. Since June 2006, he is Concordia University Re search Chair Tier I in Wireless Communications and Networking. His current research interests include machine-to-machine communications, IoT, 5G and beyond technologies, single/multiuser multiple-input multiple-output communications, space-time processing, cognitive radios, wireless networks. Dr. Hamouda served(ing) as Co-chair of the IoT and Sensor Networks Sympo sium of the GC22, TPC Co-chair of the ITC/ADC 2022 conference, track Co-Chair: Antenna Systems, Propagation, and RF Design, IEEE Vehicular Technology Conference (VTC Fall’20), Tutorial Chair of IEEE Canadian Conference in Electrical and Computer Engineering (CCECE 2020), General Co-Chair, IEEE SmartNets 2019 Conference, Co-Chair of the MAC and Cross Layer Design Track of the IEEE (WCNC) 2019, Co-chair of the Wireless Communications Symposium of the IEEE ICC18, Co-chair of the Ad-hoc, Sensor, and Mesh Networking Symposium of the IEEE GC17, Technical Co-chair of the Fifth International Conference on Selected Topics in Mobile \& Wireless Networking (MoWNet2016), Track Co-Chair: Multiple Antenna and Cooperative Communications, IEEE Vehicular Technology Conference (VTC Fall’16), Co-Chair: ACM Performance Evaluation of Wireless Ad Hoc, Sensor, and Ubiquitous Networks (ACMPE-WASUN’14) 2014, Technical Co chair of the Wireless Networks Symposium, 2012 Global Communications Conference, the Ad hoc, Sensor, and Mesh Networking Symposium of the 2010 ICC, and the 25th Queens Biennial Symposium on Communications. He also served as the Track Co-chair of the Radio Access Techniques of the 2006 IEEE VTC Fall 2006 and the Transmission Techniques of the IEEE VTC-Fall 2012. From September 2005 to November 2008, he was the Chair of the IEEE Montreal Chapter in Communications and Information Theory. He is an IEEE ComSoc Distinguished Lecturer. He received numerous awards, including the Best Paper Awards of the IEEE GC 2023, IEEE GC 2020, IEEE ICC 2021, ICCSPA 2019, IEEE WCNC 2016, IEEE ICC 2009, and the IEEE Canada Certificate of Appreciation in 2007 and 2008. He served as an Associate Editor for the IEEE COMMUNICATIONS LETTERS, IEEE TRANSACTIONS ON SIGNAL PROCESSING, IEEE COMMUNICATIONS SURVEYS ANDTUTORIALS, IET WIRELESS SENSOR SYSTEMS, IEEE WIRELESS COMMUNICATIONS LETTERS, TRANSACTIONS ON VE HICULAR TECHNOLOGY, and currently serves as an Editor for the IEEE TRANSACTIONS ON COMMUNICATIONS, IEEE TRANSACTIONS ON WIRELESS COMMUNICATIONS, IEEE IoT journal.
\end{IEEEbiography}


\begin{thebibliography}{45}

\bibitem{1}
A. Azab, M. Khasawneh, S. Alrabaee, K. K. R. Choo, and M. Sarsour, ``Network traffic classification: Techniques, datasets, and challenges,'' \textit{Digital Communications and Networks}, vol. 8, no. 2, pp. 272--287, Apr. 2022.

\bibitem{2}
M. Saqib, H. Elbiaze, and Roch H. Glitho, ``Adaptive In-Network Traffic Classifier: Bridging the Gap for Improved QoS by Minimizing Misclassification,'' \textit{IEEE Open Journal of the Communications Society}, Vol. 5 pp. 677--689, Jan 2024.

\bibitem{3}
Cisco, ``Cisco Visual Networking Index: Forecast and Methodology, 2016--2021,'' White Paper, Jun. 2017. [Online]. Available: https://www.cisco.com/c/en/us/solutions/collateral/service-provider/visual-networking-index-vni/complete-white-paper-c11-481360.html

\bibitem{4}
Y. Feng, J. Li, J. Mirkovic, C. Wu, C. Wang, H. Ren, J. Xu, and Y. Liu, ``Unmasking the Internet: A Survey of Fine-Grained Network Traffic Analysis,'' \textit{IEEE Communications Surveys \& Tutorials}, Feb. 2025.

\bibitem{5}
E. Paolini, L. Valcarenghi, L. Maggiani, and N. Andriolli, ``Real-Time Network Packet Classification Exploiting Computer Vision Architectures,'' \textit{IEEE Open Journal of the Communications Society}, Vol. 5 pp. 1155--1166, Feb. 2024.

\bibitem{6}
A. Shahraki, M. Abbasi, A. Taherkordi, and A. D. Jurcut, ``A comprehensive survey on 6G networks: Applications, core services, enabling technologies, and challenges,'' arXiv:2101.12475, Jan. 2021.

\bibitem{7}
M. Lopez-Martin, B. Carro, A. Sanchez-Esguevillas, and J. Lloret, ``Deep learning for network traffic monitoring and analysis (NTMA): A survey,'' \textit{Computer Communications}, vol. 170, pp. 19--33, Mar. 2021.

\bibitem{8}
D. Di Monda, A. Montieri, V. Persico, P. Voria, M. De Ieso, and A. Pescapè, ``Few-Shot Class-Incremental Learning for Network Intrusion Detection Systems,'' \textit{IEEE Open Journal of the Communications Society}, Vol. 5 pp. 6736--6757, Oct 2024.

\bibitem{9}
P. Khandait, N. Hubballi, and B. Mazumdar, ``Efficient keyword matching for deep packet inspection based network traffic classification,'' \textit{IEEE International Conference on Communication Systems \& Networks (COMSNETS)}, pp. 567--570, Mar. 2020.

\bibitem{10}
A. Nascita, G. Aceto, D. Ciuonzo, A. Montieri, V. Persico, and A. Pescapé, ``A Survey on Explainable Artificial Intelligence for Internet Traffic Classification and Prediction, and Intrusion Detection,'' \textit{IEEE Communications Surveys \& Tutorials}, Nov. 2024.

\bibitem{11}
R. Raveendran and R. Menon, ``An efficient method for Internet traffic classification and identification using statistical features,'' \textit{Int. Journal Eng. Res. Technol}, vol. 4, no. 7, pp. 297--303, 2014.

\bibitem{12}
G. Draper-Gil, A. Habibi, M. Saiful, and A. Ghorbani, ``Characterization of encrypted and VPN traffic using time-related features,'' \textit{Proc. 2nd Int. Conf. Inf. Syst. Security Privacy (ICISS)}, pp. 407--414, 2016.

\bibitem{13}
R. H. Serag, M. S. Abdalzaher, H. Elsayed, M. Sobh, M. Krichen, and M. Salim, ``Machine-Learning-Based Traffic Classification in Software-Defined Networks,'' \textit{Electronics}, vol. 13, no. 6, Mar. 2024.

\bibitem{14}
W. Lin and Y. Chen, ``Robust Network Traffic Classification Based on Information Bottleneck Neural Network,'' \textit{IEEE Access}, vol. 12, pp. 150169--150179, Oct. 2024.

\bibitem{15}
Y. Ma, Z. Li, H. Xue, and J. Chang, ``A balanced supervised contrastive learning-based method for encrypted network traffic classification,'' \textit{Computers \& Security Journal}, vol. 145, pp. 104023, Oct. 2024.

\bibitem{16}
X. Xiao, S. Wang, G. Hu, X. Luo, Q. Li, K. Mao, B. Zhang, and S. Xia, ``RBLJAN: Robust Byte-Label Joint Attention Network for Network Traffic Classification,'' \textit{IEEE Transactions on Dependable and Secure Computing}, vol. 22, no. 3, pp. 2161--2178, Oct. 2024.

\bibitem{17}
K. Lin, X. Xu, and Y. Jiang, ``A New Semi-supervised Approach for Network Encrypted Traffic Clustering and Classification,'' \textit{IEEE International Conference on Computer Supported Cooperative Work in Design (CSCWD)}, May. 2022.

\bibitem{18}
E. Horowicz, T. Shapira, and Y. Shavitt, ``Self-Supervised Traffic Classification: Flow Embedding and Few-Shot Solutions,'' \textit{IEEE Transactions on Network and Service Management}, vol. 21, no. 3, pp. 3054--3067, June 2024.

\bibitem{19}
X. Lin, G. Xiong, G. Gou, Z. Li, J. Shi, and J. Yu, ``ET-BERT: A Contextualized Datagram Representation with Pre-training Transformers for Encrypted Traffic Classification,'' \textit{arXiv:2202.06335}, Feb. 2022.

\bibitem{20}
R. Zhao, M. Zhan, X. Deng, Y. Wang, Y. Wang, G. Gui, and Z. Xue, ``Yet Another Traffic Classifier: A Masked Autoencoder Based Traffic Transformer with Multi-Level Flow Representation,'' \textit{Proceedings of the AAAI Conference on Artificial Intelligence}, vol. 37, no. 4, pp. 5420--5427, 2023.

\bibitem{21}
M. Towhid and N. Shahriar, ``Encrypted Network Traffic Classification using Self-supervised Learning,'' \textit{IEEE International Conference on Network Softwarization (NetSoft)}, pp. 366--375, Aug. 2022.

\bibitem{22}
Y. Heng, V. Chandrasekhar, and J. G. Andrews, ``UTMobileNetTraffic2021: A Labeled Public Network Traffic Dataset,'' \textit{IEEE Networking Letters}, vol. 3, no. 3, pp. 156--160, 2021.

\bibitem{23}
J. L. Guerra, C. Catania, and E. Veas, ``Datasets are not enough: Challenges in labeling network traffic,'' \textit{Comput. Secur.}, vol. 120, p. 102810, Jun. 2022.

\bibitem{24}
D. Soukup, P. Tisovčík, K. Hynek, and T. Čejka, ``Towards Evaluating Quality of Datasets for Network Traffic Domain,'' \textit{IEEE Int. Conf. Network and Service Management (CNSM)}, 2021.


\bibitem{25}
J. Krupski, M. Iwanowski, and W. Graniszewski, ``On the right choice of data from popular datasets for Internet traffic classification,'' \textit{Computer Communications Journal}, vol. 233, pp. 108068, March 2025.


\bibitem{26}
C. G. Northcutt, L. Jiang, and I. L. Chuang, ``Confident learning: Estimating uncertainty in dataset labels,'' \textit{Journal of Artificial Intelligence Research}, vol. 70, pp. 1373--1411, Apr. 2021.

\bibitem{27}
W. Cui, R. Hosseinzadeh, J. Ma, T. Wu, Y. Sui, and K. Golestan, ``Tabular Data Contrastive Learning via Class-Conditioned and Feature-Correlation Based Augmentation,'' \textit{	arXiv:2404.17489}, Apr. 2024.


\bibitem{28}
N. Hubballi, M. Swarnkar, and M. Conti, ``BitProb: Probabilistic Bit Signatures for Accurate Application Identification,'' \textit{IEEE Transactions on Network and Service Management}, vol. 17, no. 3, pp. 1730--1741, Sep. 2020.


\bibitem{29}
A. Salau and M. Beyene, ``Software defined networking based network traffic classification using machine learning techniques,'' \textit{Scientific Reports}, vol. 14, Aug. 2024.

\bibitem{30}
M. R. Choudhury, P. Acharjee, and A. T. George, ``Network Traffic Classification Using Supervised Learning Algorithms,'' \textit{IEEE International Conference on Computer, Electrical \& Communication Engineering (ICCECE)}, Apr. 2023.

\bibitem{31}
M. Lotfollahi, M. J. Siavoshani, R. S. H. Zade, and M. Saberian, ``Deep packet: A novel approach for encrypted traffic classification using deep learning,'' \textit{Soft Computing}, vol. 24, no. 3, pp. 1999--2012, 2020.

\bibitem{32}
A. Jenefa, S. Sam, V. Nair, B. G. Thomas, A. S. George, and R. Thomas, ``A Robust Deep Learning-based Approach for Network Traffic Classification using CNNs and RNNs,'' \textit{IEEE International Conference on Signal Processing and Communication (ICSPC)}, pp. 106--111, May. 2023.

\bibitem{33}
G. D’Angelo and F. Palmieri, ``Network traffic classification using deep convolutional recurrent autoencoder neural networks for spatial–temporal features extraction,'' \textit{Journal of Network and Computer Applications}, vol. 173, pp. 102890, Jan. 2021.

\bibitem{34}
H. Zhang, L. Yu, X. Xiao, Q. Li, F. Mercaldo, X. Luo, and Q. Liu, ``TFE-GNN: A Temporal Fusion Encoder Using Graph Neural Networks for Fine-grained Encrypted Traffic Classification,'' \textit{	arXiv:2307.16713}, Jul. 2023.

\bibitem{35}
L. Liu, Y. Yu, Y. Wu, Z. Hui, and J. Hu, ``Method for multi-task learning fusion network traffic classification to address small sample labels,'' \textit{Scientific Reports}, vol. 14, Jan. 2024.


\bibitem{36}
S. Dadkhah, H. Mahdikhani, P. K. Danso, A. Zohourian, K. A. Truong, and A. A. Ghorbani, ``Towards the development of a realistic multidimensional IoT profiling dataset,'' \textit{19th Annual International Conference on Privacy, Security \& Trust (PST2022)}, Aug. 2022.

\bibitem{37}
W. Zhang, L. Zhang, X. Zhang, Y. Wang, P. Liu, and G. Gui, ``Intelligent Unsupervised Network Traffic Classification Method Using Adversarial Training and Deep Clustering for Secure Internet of Things,'' \textit{MDPI Future Internet Journal}, vol. 15, no. 9, pp. 298--318, 2023.

\bibitem{38}
B. Gao, Y. Yang, Z. Gao, P. Yu, R. Lyu, and S. Chen, ``Unsupervised Network Traffic Classification Based on Multi-Source Synergistic Distribution Alignment,'' \textit{GLOBECOM 2023 - 2023 IEEE Global Communications Conference}, Feb. 2024.

\bibitem{39}
Yang, R., Yu, A., Cai, L., and Meng, D., ``Subspace clustering via graph auto-encoder network for unknown encrypted traffic recognition,'' \textit{Cybersecurity}, vol. 5, no. 29, 2022.


\bibitem{40}
Y. Huo, C. Song, M. Zhou, R. Lv, and Y. Yang, ``A Novel Approach for Semi-Supervised Network Traffic Classification,'' \textit{IEEE International Conference on Advanced Infocomm Technology (ICAIT)}, pp. 64--70, Aug. 2022.

\bibitem{41}
M. A. Aleisa, ``Traffic classification in SDN-based IoT network using two-level fused network with self-adaptive manta ray foraging,'' \textit{Scientific Reports}, vol. 15, Jan. 2025.

\bibitem{42}
S. Xu, J. Han, Y. Liu, H. Liu, and Y. Bai, ``Few-shot traffic classification based on autoencoder and deep graph convolutional networks,'' \textit{Scientific Reports}, vol. 15, Mar. 2025.

\bibitem{43}
J. Luxemburk, K. Hynek, T. Cejka, A. Lukacovic, and P. Šiška, ``CESNET-QUIC22: A large one-month QUIC network traffic dataset from backbone lines,'' \textit{Data in Brief}, vol. 46, Feb. 2023.

\bibitem{44}
S. Rezaei, and X. Liu, ``How to Achieve High Classification Accuracy with Just a Few Labels: A Semi-supervised Approach Using Sampled Packets,'' \textit{	arXiv:1812.09761}, May 2020.

\bibitem{45}
H. Zhang, L. Yu, X. Xiao, Q. Li, F. Mercaldo, X. Luo, and Q. Liu, ``TFE-GNN: A Temporal Fusion Encoder Using Graph Neural Networks for Fine-grained Encrypted Traffic Classification,'' \textit{	arXiv:2307.16713 }, Jul. 2023.

\end{thebibliography}
\end{document}